\documentclass[aps,prl,twocolumn,showpacs,superscriptaddress,groupedaddres,nofootinbib,preprintnumbers]{revtex4}  % for review and submission
\usepackage{graphicx}  % needed for figures
\usepackage{dcolumn}   % needed for some tables
\usepackage{bm,amsmath}        % for math
\usepackage{amssymb}   % for math
\usepackage{gensymb}
\usepackage{enumerate}
\usepackage{pifont}
\usepackage{color}

\newcommand{\edr}[1]{} %Edit: removal

\newcommand{\amc}{{\sc MadGraph5\textunderscore}a{\sc MC@NLO}}
\newcommand{\beq}{\begin{equation}} 
\newcommand{\eeq}{\end{equation}} 
\newcommand{\bea}{\begin{eqnarray}} 
\newcommand{\eea}{\end{eqnarray}}

\begin{document}

\title{Measuring the mass, width, and couplings of semi-invisible resonances\\ with the Matrix Element Method}
\author{Amalia Betancur} 
\affiliation{Grupo F\' isica Te\' orica y Aplicada, Universidad EIA, A.A. 7516, Medell\' in, Colombia}
\affiliation{Instituto de F\' isica, Universidad de Antioquia, Calle 70 No. 52-21, Medell\' in, Colombia}
\author{Dipsikha Debnath} \affiliation{Physics Department, University
  of Florida, Gainesville, FL 32611, USA}
\author{James~S.~Gainer} \affiliation{Dept.~of Physics and Astronomy, 
University of Hawaii, Honolulu, HI 96822, USA}
\author{Konstantin~T.~Matchev} \affiliation{Physics Department,
  University of Florida, Gainesville, FL 32611, USA}
\author{Prasanth Shyamsundar} \affiliation{Physics Department, University
  of Florida, Gainesville, FL 32611, USA}
\date{April 29, 2019}

\begin{abstract}
We demonstrate the use of the Matrix Element Method (MEM) for the measurement of masses, widths, and couplings 
in the case of single or pair production of semi-invisibly decaying resonances. For definiteness, we consider the two-body 
decay of a generic resonance to a visible particle from the Standard Model (SM) and a massive invisible particle. It is well
known that the mass difference can be extracted from the endpoint of a transverse kinematic variable like 
the transverse mass, $M_T$, or the Cambridge $M_{T2}$ variable, but measuring the overall mass scale is a very difficult problem.
We show that the MEM can be used to obtain not only the absolute mass scale, but also the width of the resonance and the 
tensor structure of its couplings. Apart from new physics searches, our results can be readily applied to the case of SM $W$ boson 
production at the CERN Large Hadron Collider (LHC), where one can repeat the measurements of the $W$ properties 
in a general and model-independent framework.
\end{abstract}

\pacs{13.85.Hd, %inelastic scattering: many particle final states 
14.70.Fm,   %W bosons
14.80.-j   % other particles (including hypothetical)
}

\preprint{UH-511-1280-2017}

\maketitle
\begin{figure}[t]
\includegraphics[width=0.49\columnwidth]{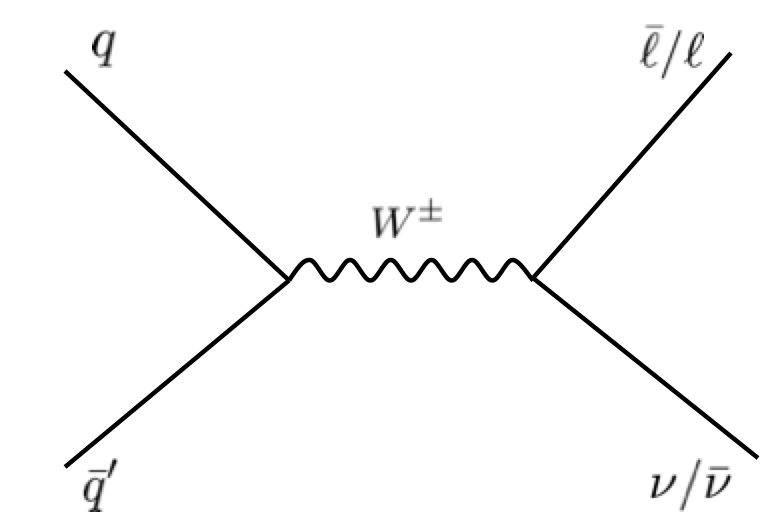} 
\includegraphics[width=0.49\columnwidth]{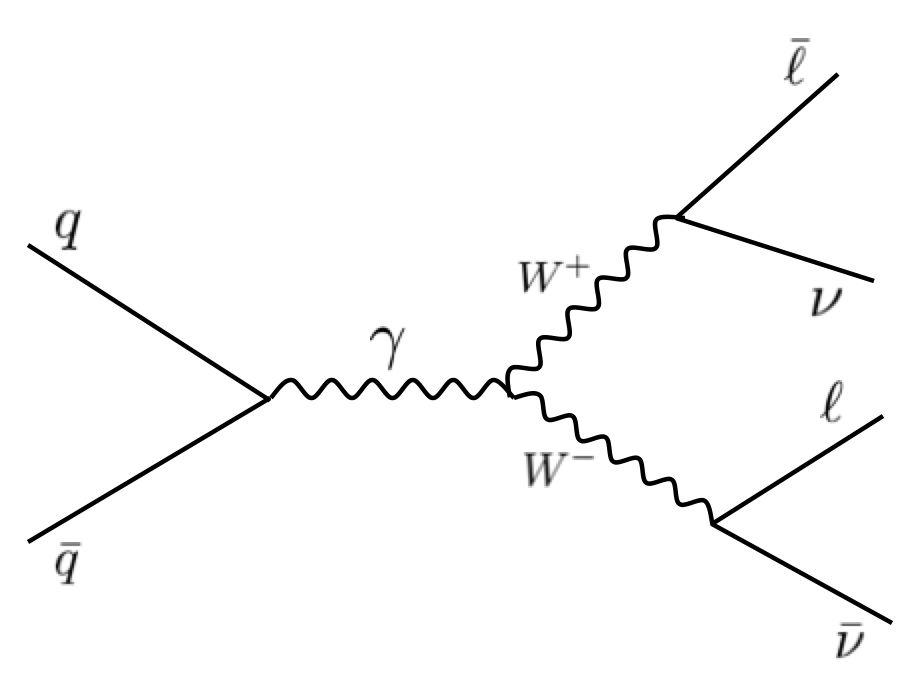} 
(a) \hskip 3.8cm (b)
\caption{\label{fig:diagrams}
The event topologies considered in this paper: (a) single and (b) pair 
production of a $W$-like resonance decaying leptonically.
}
\end{figure}

{\bf Introduction.}~~ 
The dark matter problem is the biggest mystery in particle physics today~\cite{DMreview}.
It greatly motivates the current experimental efforts to discover new physics beyond the 
Standard Model (BSM) at the LHC.
In typical BSM models, dark matter particles are produced at the LHC as the end products
of the cascade decays of heavier particles. Generally speaking, the longer the cascade, the 
more handles we have at our disposal to measure particle properties like masses, widths, couplings, etc.
Hence, the most challenging cases are actually the simplest ones, illustrated in Fig.~\ref{fig:diagrams} for the case of
single production (diagram (a)) and pair-production (diagram (b)).  For these processes,
the classic method of kinematic endpoints~\cite{Hinchliffe:1996iu} fails to 
determine the complete mass spectrum --- there is only one observable endpoint\footnote{In order 
to preform this measurement in practice, one can use any one of several variables, e.g., the 
transverse mass, $M_T$, of the parent particle \cite{Smith:1983aa,Barger:1983wf}, or the 
transverse momentum, $p_T$, of the visible daughter particle for the case of Fig.~\ref{fig:diagrams}(a),
or the Cambridge $M_{T2}$ variable \cite{Lester:1999tx,Barr:2003rg}, the 
contransverse mass $M_{CT}$ \cite{Tovey:2008ui}, or their 1D variants 
\cite{Konar:2009wn,Matchev:2009ad} for the case of Fig.~\ref{fig:diagrams}(b). },
and the best one can do is to obtain a relationship\footnote{See eq.~(\ref{MTendpoint}) below.} 
$M_{W}(M_{\nu})$ between the mass of the parent, $M_W$, and the mass of the daughter particle, $M_\nu$, 
leaving the overall mass scale undetermined. 
In particular,  standard methods for measuring the $W$ boson mass, 
such as via properties of the transverse mass spectra like
 the Jacobian peak~\cite{Smith:1983aa, Barger:1983wf},   
or by determining $M_W/M_Z$ by comparison with $Z$ boson
events~\cite{Giele:1998uh}, are not sufficient in this scenario.

%%%%%%%%%%%%% Beginning OF FIGURE ################%%%%%%%%%%%%
%%%%%%%%%%%%% End OF FIGURE %%%%%%%%%%%%%%%%%%%%%%%%%%%%%%%%%

One possibility that has been suggested in the literature for determining the overall mass scale is to go
beyond the leading order diagrams of Fig.~\ref{fig:diagrams} and consider hard initial state radiation (ISR),
which provides a kick to the system in the transverse plane. In the presence of ISR, the functional dependence
$M_{W}(M_{\nu})$, derived from either $M_T$ for the case of Fig.~\ref{fig:diagrams}(a) \cite{Gripaios:2007is,Barr:2007hy}
or from $M_{T2}$ for the case of Fig.~\ref{fig:diagrams}(b) \cite{Cho:2007qv,Cho:2007dh,Burns:2008va,Matchev:2009fh,Barr:2009jv,Konar:2009wn,Konar:2009qr},
exhibits a kink at the true value, $M_{\nu}^{true}$, of the daughter particle. However, the kink is not very visible 
unless we demand very hard ISR, which causes a significant loss in statistics. 

Given the difficulty of measuring the overall mass scale, it stands to reason that one should 
use the MEM~\cite{kondo1,kondo2,kondo3,dalitz,oai:arXiv.org:hep-ex/9808029,vigil,canelli,abazov,Gainer:2013iya}, which, like similar 
multivariate analyses (MVA)~\cite{MVA},
has been used with success in LHC experiments.  A particularly
dramatic example has been the use of the MEM in the four-lepton
channel for the discovery of the Higgs Boson~\cite{Higgs Discovery}
and the measurement of its properties~\cite{Chatrchyan:2012jja,Aad:2013xqa,Chatrchyan:2013mxa}.  Such
analyses used variables, such as MELA KD~\cite{Gao:2010qx,DeRujula:2010ys,Bolognesi:2012mm} or
MEKD~\cite{Avery:2012um,Chen:2013waa} that involve the ratio of signal and background
matrix elements\footnote{Strictly speaking, those analyses were using MEM-inspired optimal variables, 
whose distributions were used later to form the likelihood in a template method instead of as in eq.~(\ref{eq:likelihood}).}.  
Clearly these variables are, therefore, optimized to the appropriate signal and background hypotheses.  

To our knowledge, the MEM has thus far been underutilized for the purposes of mass spectrum measurements
of the sort we describe here. This is perhaps due to the practical challenges one usually encounters in the implementation of the MEM, e.g.,
the presence of instrumental and/or reducible physics backgrounds;
the need to account for effects like the finite detector resolution, the underlying event, jet fragmentation;
the challenge of integrating peaked integrand structures in phase space;
incorporating higher order corrections to the matrix element, etc.
\cite{Gainer:2013iya}.
Despite these practical limitations, the physical transparency of the MEM greatly motivates expanding its role in LHC data analyses.
Previously, Ref.~\cite{Alwall:2009sv} applied the MEM to the case of squark pair production 
in SUSY (which has the same event topology as Fig.~\ref{fig:diagrams}(b)) and demonstrated that one can 
not only recover the endpoint measurement information, but also place a restriction on the allowed overall mass scale. 
Similar conclusions were drawn in Ref.~\cite{Artoisenet:2010cn} for the example of smuon pair production as in Fig.~\ref{fig:diagrams}(b). 
We would like to revisit these earlier studies and extend the application of the MEM to the measurement of the 
remaining properties of the particles involved, namely the width and the chirality of the couplings.
(The discrete choices of the spin assignments of the particles in the decay chain were previously
investigated in Ref.~\cite{Chen:2010ek}.)

{\bf Formulation of the problem.}~~We begin by considering single production of a resonance, 
$W^+$, which decays semi-invisibly via a two-body decay into a visible SM particle, $\bar{\ell}$, 
and an invisible particle, $\nu$. For definiteness, we shall take $\bar{\ell}$ to be an anti-lepton 
(positron or antimuon) and $\nu$ to be an invisible particle, which can be a SM neutrino or
some BSM dark matter candidate. The $W^+$ resonance can be produced singly,
as shown in Fig.~\ref{fig:diagrams}(a), or as part of a $W^+ W^-$ pair, as in the $s$-channel\footnote{We focus 
on the photon-mediated $s$-channel diagram for simplicity: 
since the $W$ is charged, it must couple to photons, so that the diagram 
of Fig.~\ref{fig:diagrams}(b) is guaranteed to exist. In principle, there can be 
additional $t$- and $u$-channel pair-production diagrams, but this requires 
that the $W$ couples to quarks as well, in which case the single production from
Fig.~\ref{fig:diagrams}(a) should dominate. There could also be $s$-channel diagrams mediated by $Z$ or other more exotic gauge bosons, 
but this possibility also involves additional assumptions. All of those complications can be easily incorporated in the analysis, and the 
MEM would still work, but our discussion would become more opaque.} 
diagram of Fig.~\ref{fig:diagrams}(b). As suggested by our notation, this setup includes, 
but is not limited to, the SM production of $W^+$ bosons decaying leptonically. 
In particular, the process of Fig.~\ref{fig:diagrams}(a) may refer to 
the production of a charged Higgs scalar \cite{Altmannshofer:2016zrn},
a charged slepton in supersymmetry (SUSY) models with $R$-parity violation \cite{Kalinowski:1997zt,Hewett:1998fu}, 
or a new $W'$ heavy gauge boson \cite{Khachatryan:2016jww,Aaboud:2017efa}. 
Similarly, the process of Fig.~\ref{fig:diagrams}(b) may be interpreted as the pair-production of (perhaps quarkophobic) charged Higgs bosons
\cite{Eichten:1984eu,Willenbrock:1986ry} or $W'$ bosons \cite{Belyaev:2006jh}, of
charginos \cite{Barbieri:1991vk},  Kaluza-Klein leptons \cite{Cheng:2002ab,Barr:2005dz}, or
sleptons \cite{delAguila:1990yw,Baer:1993ew}.
However, for definiteness, in our simulations below we shall assume that the $W^\pm$ are spin-$1$ particles while $\ell (\bar{\ell})$ and $\nu (\bar{\nu})$ are spin $1/2$, as in
the SM. We shall parametrize the $W$ couplings to leptons (quarks) as 
$g^\ell_RP_R+g^\ell_LP_L$ ($g^q_RP_R+g^q_LP_L$), where $P_{L,R}=(1\mp\gamma_5)/2$ are the usual chiral projectors.

The (normalized) kinematic distributions of the leptons in the final state will depend on five model parameters:
\beq
\left\{ M_W, M_\nu, \Gamma_W, \varphi_q, \varphi_\ell \right\},
\label{parameters}
\eeq
where $M_W$ ($M_\nu$) is the mass of the parent (daughter) particle, $\Gamma_W$ is the width of the parent,
and 
\beq
\tan\varphi_q\equiv \frac{g^q_R}{g^q_L},  \quad \tan\varphi_\ell\equiv \frac{g^\ell_R}{g^\ell_L},
\label{chirality}
\eeq
so that the angles $\varphi_q$ and $\varphi_\ell$ encode the information about the chirality of the $W$
couplings to quarks and leptons, respectively.

Given this general setup, our main goal in this paper will be to attempt to measure {\em all five} of the model parameters in
eq.~(\ref{parameters}). As already discussed in the introduction, this is not a trivial task. Measuring the mass splitting 
is relatively straightforward (see Fig.~\ref{fig:ptzero} below), but fixing the remaining four parameters requires subtle
measurements of the relevant kinematic distributions. We shall make use of the MEM, which is ideally suited for our purposes. 
Along the way we shall also study the dependence of the relevant kinematic variables on 
the underlying parameters (\ref{parameters}), highlighting the cases when a certain variable depends strongly on 
a particular parameter. This not only provides intuitive understanding of our 
main results (namely, Figs.~\ref{fig:simul}, \ref{fig:GwMwPairProd} and \ref{fig:chirality2} below)
which are obtained with the MEM, but also offers an alternative approach to estimate the parameters
(\ref{parameters}) by template-fitting to the corresponding sensitive kinematic variables.

{\bf The Matrix Element Method.}~~
In the MEM, the likelihood for a given event with a measured set of $N_f^{vis}$ visible final state 4-momenta $\{P_j^{\,vis}\}$, $j=3,...,N_f^{vis} + 2$, is defined as 
\begin{eqnarray} \label{eq:likelihood}
{\cal P}(\{P_j^{\,vis}\} |\alpha) &=& \frac{1}{\sigma_{\alpha}}
\biggr[ \prod_{j=1}^{N_{f}} \int \frac{d^3 p_j}{(2\pi)^3 2E_j} \biggr]
W(\{P_j^{\,vis}\}, \{p_j^{vis}\} ) \nonumber \\
&\times&  
\sum_{a,b}  \frac{f_a(x_1)f_b(x_2)}{2sx_1x_2} |{\cal M}_{\alpha}(\{p_i\},\{p_j\})|^2 \nonumber\\
&\times& (2 \pi)^4 \delta^4\left(\sum_{i=1}^{2} p_i-\sum_{j = 3}^{N_f + 2} p_j\right).
\end{eqnarray}
Here $f_a$ and $f_b$ are the parton distribution functions (pdf) of the initial state partons $a$ and $b$ 
as a function of their momentum fractions $x_1$ and $x_2$, while $\sqrt{s}$ is the center of mass energy of the collider. 
${\mathcal M}_{\alpha}$ represents the theoretical matrix element for the parton-level scattering process $\{p_i\}\to\{p_j\}$ 
under a given hypothesis $\alpha$\footnote{
In general ${\mathcal M}_{\alpha}$ includes the appropriate color factor, appropriate
symmetry factors, etc. for the process under consideration, though when only one process
contributes, these factors will cancel in the ratio with the total cross section.
}, while $\sigma_{\alpha}$ is the total cross-section for the corresponding 
hypothesis after acceptances, efficiencies, etc., whose role is to normalize the total probability.
The transfer function $W(\{P_j^{vis}\},\{ p_j^{vis}\})$ incorporates all detector effects and efficiencies
by mapping the $3 N_f^{vis}$  {\em true} 4-momenta $\{p_j^{\,vis}\}$ of the final state visible particles onto the set of 
measured 4-momenta $\{P_j^{\,vis}\}$ \cite{Gainer:2014bta}\footnote{
We will use $P$ to refer to measured momentum variables and $p$ to label truth
momentum variables throughout this work.
}. In general, the final state will also contain a certain number, 
$N_f^{inv}$, of {\em invisible} particles, e.g., neutrinos or dark matter candidates.
Since their momenta are not measured, they are simply integrated over, so that in (\ref{eq:likelihood})
there are a total of $3 N_f= 3(N_f^{vis}+N_f^{inv})$ integrations.
The delta function factor on the third line of (\ref{eq:likelihood}) ensures energy-momentum conservation.

The likelihood for a set of $N$ events is simply the product of the individual likelihoods for each event $n$:
\begin{equation}
\mathcal L_\alpha = \prod_n^N {\cal P}(\{ P_j^{\,vis} \}_n|\alpha);
\end{equation}
when necessary an overall Poisson factor for the number of events can be used
to construct the so-called ``extended likelihood''.

{\bf Single $W$ production.}~~We first consider single $W$ production (Fig.~\ref{fig:diagrams}(a)).
The spin and color averaged squared matrix element for the process
$u \bar{d}\to W^+\to \bar{\ell} \nu$ is given by
\begin{eqnarray} \label{eq:matrix_W}
& ~ & \langle |\mathcal{M}|^2 \rangle = \frac{4|V_{ud}|^2}{3[(\hat{s}-M_W^2)^2 + (\Gamma_W M_W)^2]} \nonumber \\
&\times& \left[ \{ (g^{q}_{L})^2 (g^{\ell}_{L})^2 + (g^{q}_{R})^2 (g^{\ell}_{R})^2  \}
(p_{u}.p_{\overline{\ell}})(p_{\overline{d}}.p_{\nu}) \right. \nonumber\\
&&+ \left. \{ (g^{q}_{R})^2 (g^{\ell}_{L})^2+
(g^{q}_{L})^2 (g^{\ell}_{R})^2 \} (p_{\overline{d}}.p_{\overline{\ell}})(p_{u}.p_{\nu})\right]  ,
\end{eqnarray}
where the parton-level center-of-mass energy squared is $\hat{s}=(p_u+p_{\bar{d}})^2$, and 
we have generalized to the case of arbitrary fermion couplings to the $W$-like intermediate resonance.
$V_{ud}$ is the analogue of the CKM matrix element, and {\it is} the SM CKM matrix
element if we are considering production and decay of the SM $W$ boson.

If we take $p_1$ to be the momentum of the incident parton with positive $z$ momentum,
$p_2$ to be the momentum of the incident parton with negative $z$ momentum, and define
\begin{eqnarray}
F_1 &=& f_u(x_1) f_{\bar{d}}(x_2), \nonumber \\ 
F_2 &=& f_u(x_2) f_{\bar{d}}(x_1), \\ \nonumber
k_1 &=& (p_1 \cdot p_{\overline{\ell}}) (p_2 \cdot p_{\nu}), \\ \nonumber
k_2 &=& (p_1 \cdot p_{\nu}) (p_2 \cdot p_{\overline{\ell}}),  
\end{eqnarray}
we find that the likelihood for a particular event is proportional to
\begin{eqnarray}
\label{eq:magic-likelihood}
(F_1 + F_2)(k_1 + k_2) + \\ \nonumber
\cos{2\varphi_\ell} \cos{2\varphi_q} (F_1 - F_2) (k_1 - k_2).  
\end{eqnarray}
We see that only the second term depends on the helicity of the couplings.  As expected,
this term will not contribute in the absence of a longitudinal boost (when $F_1 = F_2$)
or if the lepton is emitted perpindicular to the beamline in the rest frame of the $W$ (when 
$k_1 = k_2$).  We also note that $k_1$ and $k_2$ depend only on $p_{\ell z}$, not on 
$p_{\ell T}$, so when determining the $p_{\ell T}$ distribution we get no contribution from
the second term, as the contribution from this term from each point with a given value of 
$(p_{\ell T}, p_{\ell z})$ is cancelled by the contribution of the term with $(p_{\ell T}, -p_{\ell z})$.
So it will be the $P_{\ell z}$ distribution rather than the $P_{\ell T}$ distribution that will give
us sensitivity to the chirality of couplings, as we will see in more detail below. 

In our subsequent analyses we generate events with \amc \cite{MadGraph} version $2.5.5$ 
for the parameter point with 
$M_W =  1000$ GeV, 
$M_\nu = 500$ GeV, and 
$\Gamma_W = 50$ GeV  at $\sqrt{s}=13$ TeV
without applying selection criteria (cuts) or detector simulation to the events.   
We  use {\sc MadWeight5} \cite{Artoisenet:2008zz} for computation of the weights in MEM calculations
using $\delta$-function transfer functions,
and have verified that the MadWeight results can be reproduced using (\ref{eq:matrix_W}), where appropriate.

We note that there are two relevant observables: the transverse momentum $P_{\ell T}$
and the longitudinal momentum $P_{\ell z}$ of the lepton, as the only visible particle in the final state is the
lepton, with fixed (zero) mass.  The third momentum degree of freedom corresponds to an azimuthal angle, 
which cannot have a non-trivial distribution in the absence of some very unexpected physics (or detector effects)
breaking the azimuthal symmetry.

{\bf Measurement of the mass ``difference".}~~
One quantity, related to the difference of the squared masses of the $W$ and the $\nu$, 
can be easily measured from the endpoint of the distribution of the $W$ transverse mass $M_{WT}$.
In the absence of ISR, this quantity can also be measured from the kinematic endpoint of the
 $P_{\ell T}$ distribution, $\mu$, as
\beq
\frac{M_W^2-M_\nu^2}{2 M_W}= {\rm constant} \equiv \mu.
\label{MTendpoint}
\eeq
We note that $\mu$,  the maximum value of lepton $p_T$, is also the 3-momentum of the lepton in the center of mass (CM) frame (still in the absence of ISR).
Eq.~(\ref{MTendpoint}) allows us to fix one of $M_W$ and $M_\nu$ once we have measured the other.  Thus,
at least for the time being, we shall focus on measuring the orthogonal mass degree of freedom, i.e., the overall mass scale,
by choosing the test masses to satisfy the relation (\ref{MTendpoint}).\footnote{We postpone 
the question of the simultaneous determination of both $M_W$ and $M_\nu$ until the analysis presented in Fig.~\ref{fig:simul} below.}
In other words, we shall vary one of the two masses, e.g., $M_\nu$, and then compute the other mass from eq.~(\ref{MTendpoint}) as
\beq
M_W=\mu +\sqrt{\mu^2+M_\nu ^2}.
\label{MWformula}
\eeq
%%%%%%%%%%%%% Beginning OF FIGURE ################%%%%%%%%%%%%
\begin{figure}[t]
\includegraphics[width=0.98\columnwidth]{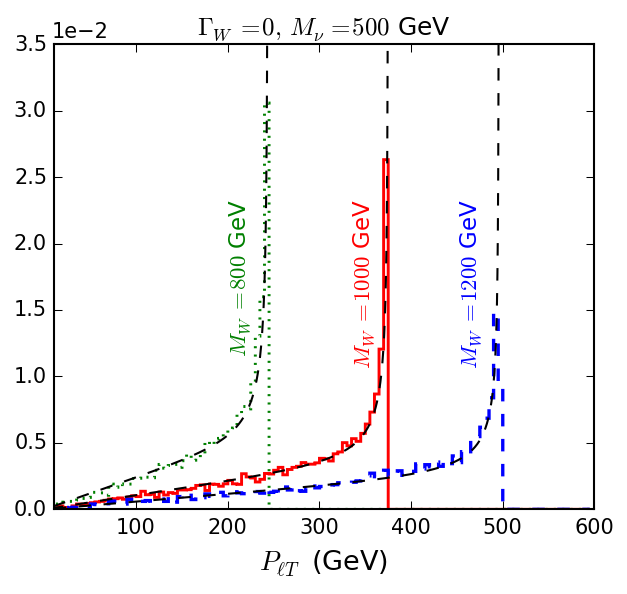}\\
\caption{\label{fig:ptzero}
Unit-normalized lepton $P_T$ distributions for single $W$ production in the limit of $\Gamma_W=0$.
The invisible particle mass is $M_\nu=500$ GeV; the mass of the parent particle is varied as shown.}
\end{figure}
%%%%%%%%%%%%% End OF FIGURE %%%%%%%%%%%%%%%%%%%%%%%%%%%%%%%%%

In Fig.~\ref{fig:ptzero}, we show the lepton $P_T$ distributions in single $W$ production for different $W$ masses; $M_W=800$ GeV (green dotted), 
$M_W=1000$ GeV (red solid), and $M_W=1200$ GeV (blue dashed) in the $\Gamma_W=0$ limit.
The mass of the invisible particle is set to $M_\nu=500$ GeV. Here we consider only left-handed fermionic couplings to the $W$ boson. 
The black dashed line gives the theoretical prediction for the {\it true} lepton $p_T$ distribution (in the absence of cuts) which is given by
\beq
\frac{1}{\sigma}\frac{d\sigma}{dp_{\ell T}} = \frac{3}{4+3\rho}\, \frac{p_{\ell T}}{\mu \sqrt{\mu ^2-p_{\ell T}^2}} 
\left( 2 -\frac{p_{\ell T}^2}{\mu^2}
+\rho
\right),
\label{dsdpt}
\eeq
where $\rho\equiv 2M_\nu^2/(M_W^2-M_\nu^2)$.
Therefore when the lepton $p_T$ reaches its maximum value, $\mu$, the longitudinal momentum $p_z$ goes to zero, and we obtain the well known
Jacobian peak in the distribution (\ref{dsdpt}).
In principle, this equation indicates that the $p_T$ spectrum depends on both $M_W$ and $M_\nu$ via the quantity, $\rho$, and the endpoint, $\mu$.  
However in practice, the dependence on $\rho$ tends to be subtle, so in practical situations we may only be able to measure $\mu$, and 
of course, a given value of $\mu$ corresponds to any $M_W$ and $M_\nu$ satisfying  eq.~(\ref{MTendpoint}).
The main point of this paper is that in addition to measuring $\mu$ (from a kinematic endpoint), by utilizing the MEM, 
we can simultaneously also obtain
(a) the mass scale, i.e., $M_W$ itself; (b) the width $\Gamma_W$; and (c) the chirality of the couplings (\ref{chirality}).

\begin{figure}[t]
\includegraphics[width=0.9\columnwidth]{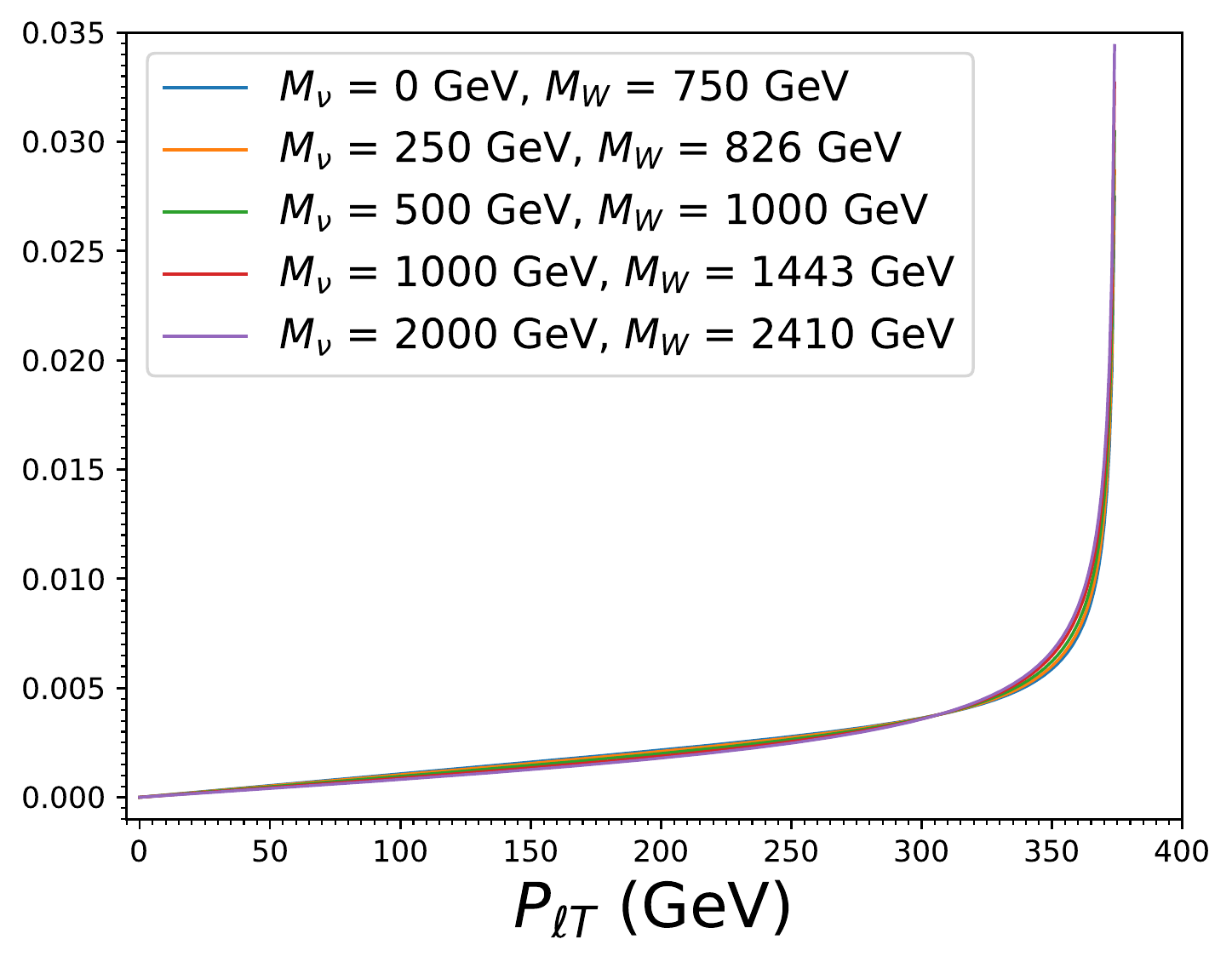}\\
\includegraphics[width=0.9\columnwidth]{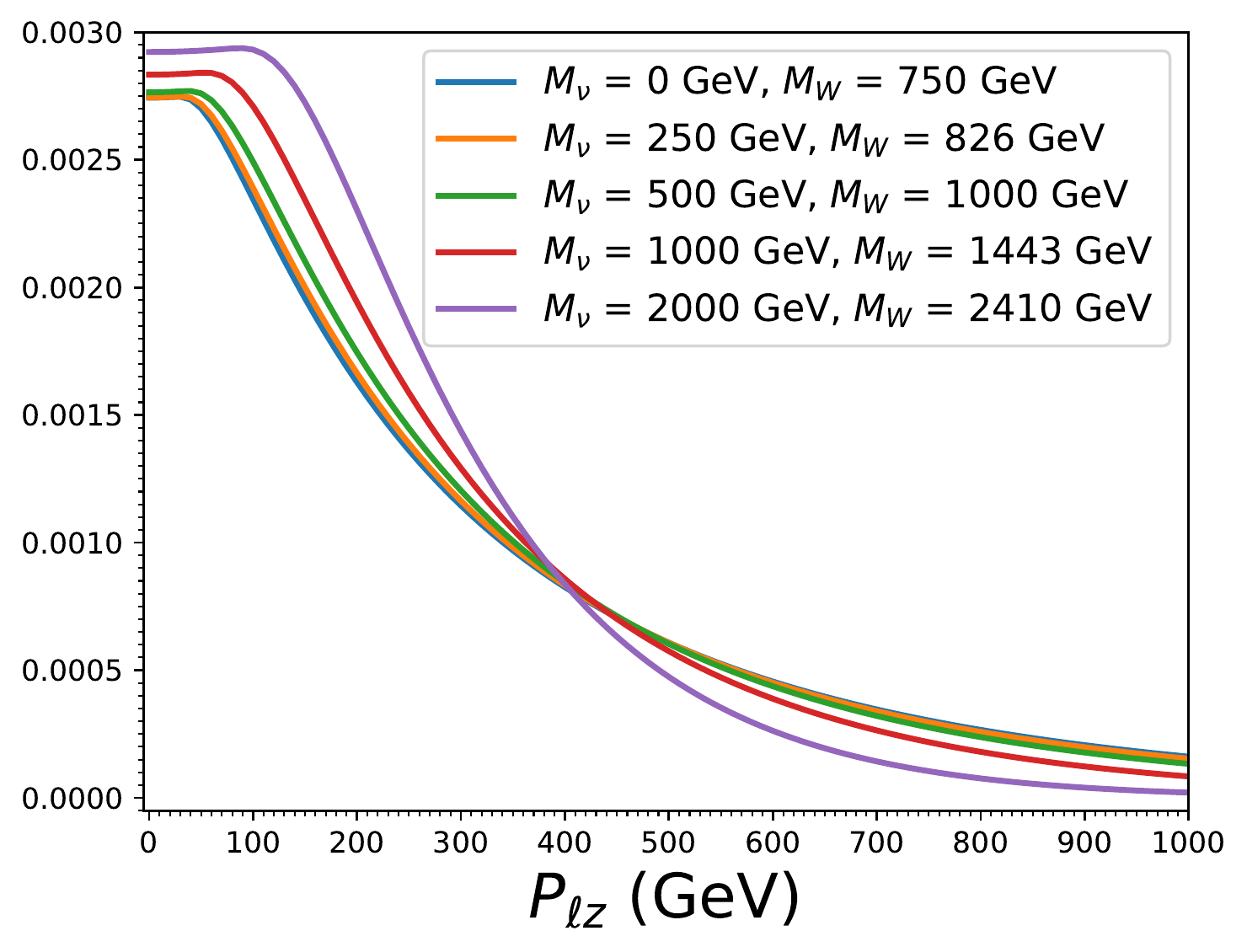}
\caption{\label{fig:pz}
Unit-normalized distributions of the lepton transverse momentum $P_T$ (top) and longitudinal momentum $P_z$ (bottom),
for different values of $M_W$ and $\Gamma_W=0$ obtained using analytical expressions and, in the case of $P_z$,
pdfs from {\sc LHAPDF}~\cite{LHAPDF}.  The mass, $M_\nu$, of the invisible particle has been fixed from the measurement
(\ref{MTendpoint}).
}
\end{figure}

{\bf Measurement of the mass scale $M_W$.}~~
The mass scale is notoriously difficult to measure; even in this very simple topology it cannot be determined from kinematic endpoint measurements alone
(unless we require hard ISR). Instead we have to rely on subtle effects. The two tools at our disposal are the distributions of the measured
lepton $P_{\ell T}$ and $P_{\ell z}$. Interestingly, the shapes of both of these distributions encode information about the mass scale, as illustrated in Fig.~\ref{fig:pz}. 
%%%%%%%%%%%%% Beginning OF FIGURE ################%%%%%%%%%%%%
%%%%%%%%%%%%% End OF FIGURE %%%%%%%%%%%%%%%%%%%%%%%%%%%%%%%%%
As seen in the top panel, the $P_T$ distribution is rather weakly sensitive to the mass scale.
However, the lower panel in Fig.~\ref{fig:pz} shows that the longitudinal momentum does contain information 
about the mass scale which can potentially be observed. 

\begin{figure}[ht]
\includegraphics[width=0.49\columnwidth]{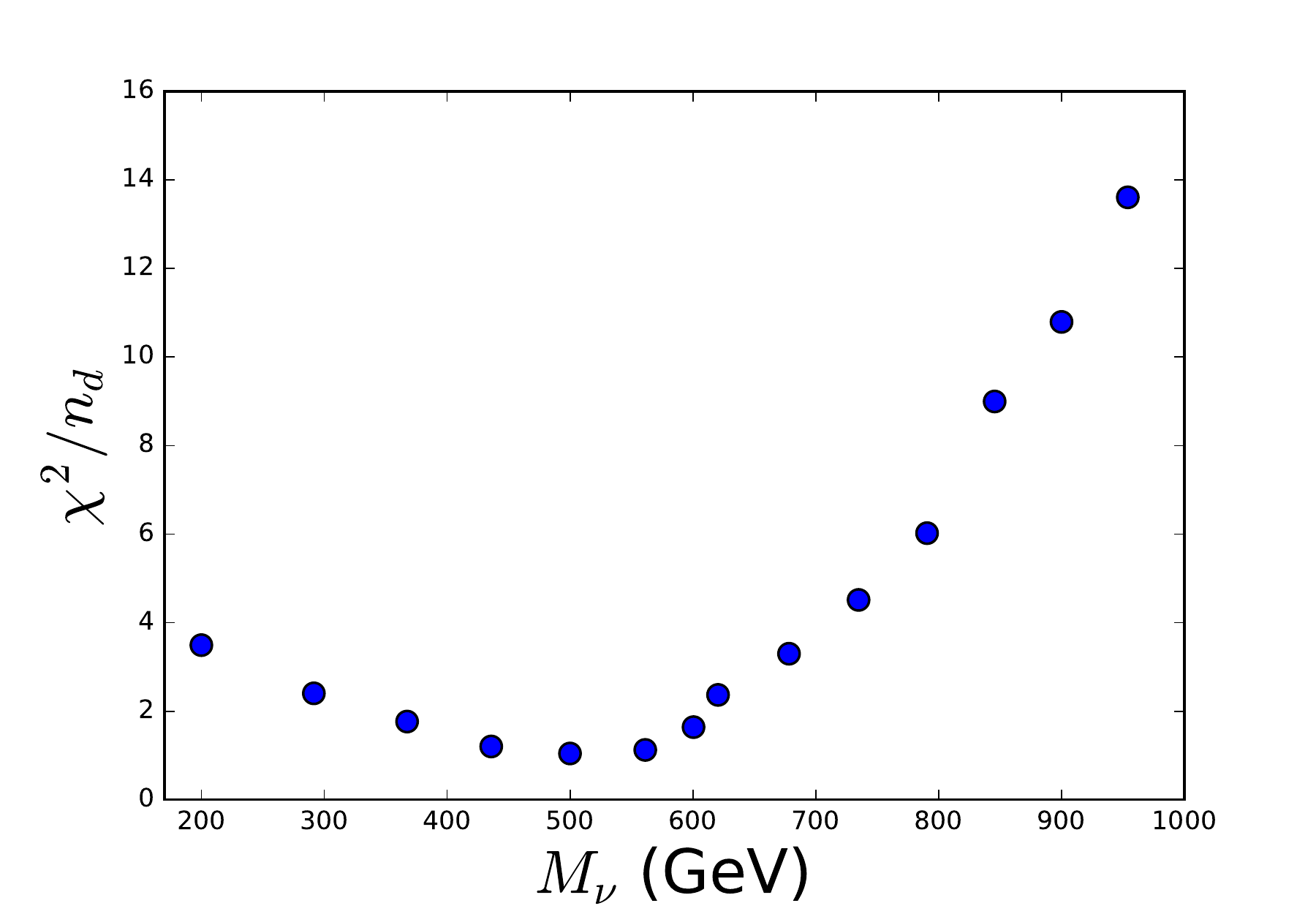}
\includegraphics[width=0.49\columnwidth]{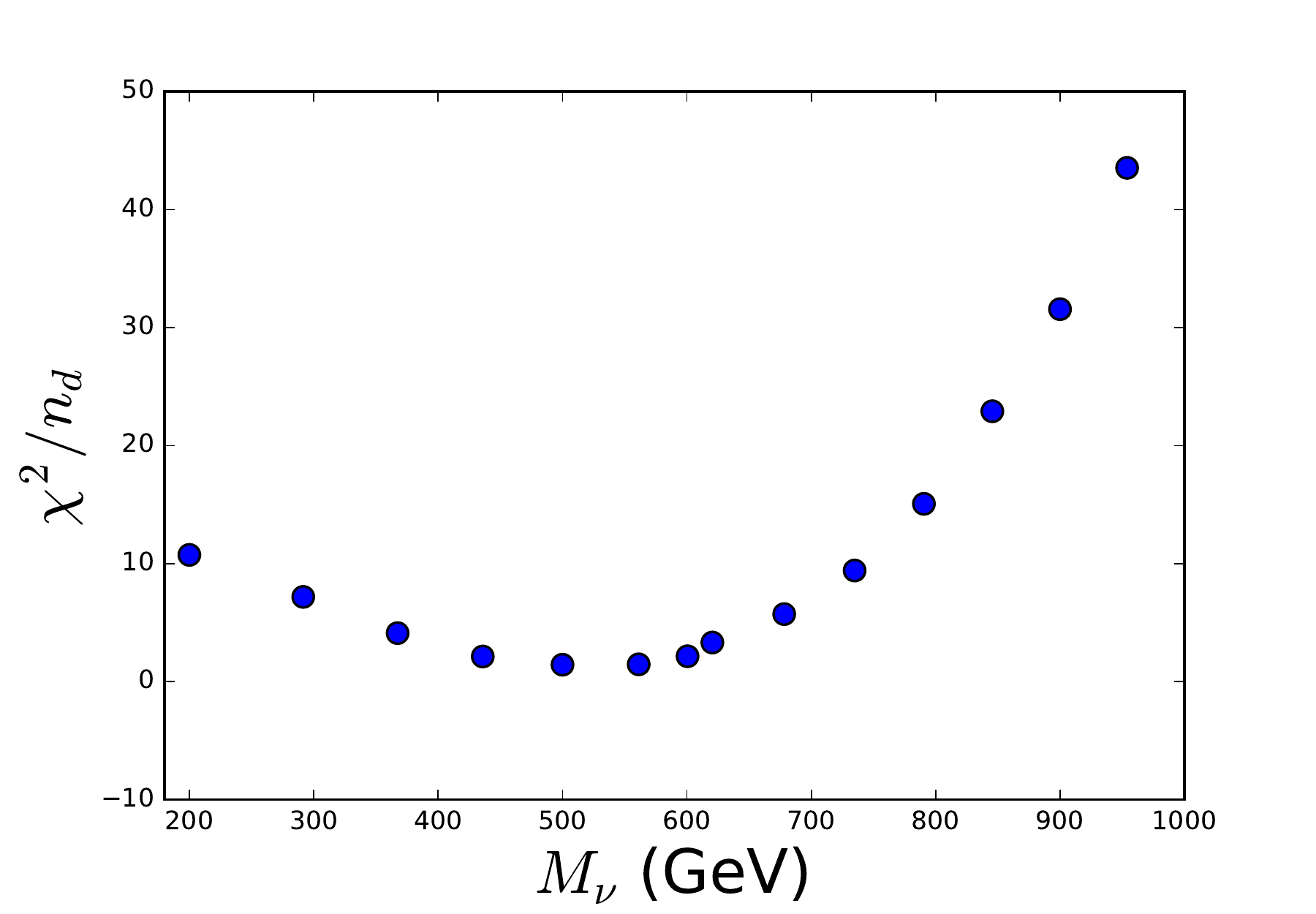}
\caption{\label{fig:template}
$\chi^2/d.o.f.$ fit to $M_\nu$ from $P_{\ell z}$ templates in $W^+$ production (left) and $W^-$ production (right).}
\end{figure}

As a proof of principle, we perform a preliminary toy exercise to find the mass scale by simply fitting to 
$P_{\ell z}$ templates generated for different mass spectra obeying the relation (\ref{MTendpoint}).
The advantage of the template method is that it avoids the time consuming integrations over the invisible momenta
that are needed for the MEM.
The result is shown in Fig.~\ref{fig:template}, where we plot the $\chi^2/d.o.f.$ for several hypothesized values of $M_\nu$
(with $M_W$ calculated from (\ref{MWformula})).
%%%%%%%%%%%%% Beginning OF FIGURE ################%%%%%%%%%%%%
%%%%%%%%%%%%% End OF FIGURE %%%%%%%%%%%%%%%%%%%%%%%%%%%%%%%%%
The right panel of Fig.~\ref{fig:template} shows results from the same exercise, but for the 
case of $d\bar{u}\to W^-\to \ell\bar{\nu}$.

The minima of the  $\chi^2$ curves in Fig.~\ref{fig:template} are near the true value, $M_\nu = 500$ GeV;
this suggests that the template method works in principle. However, the MEM will be more sensitive, as it (1) uses the
correlations among $P_{\ell T}$ and $P_{\ell z}$ in the data and (2) incorporates the dependence on the remaining parameters in eq.~(\ref{parameters}),
which makes it possible to do a {\em simultaneous} measurement of several parameters. 

\begin{figure}[t]
\includegraphics[width=0.98\columnwidth]{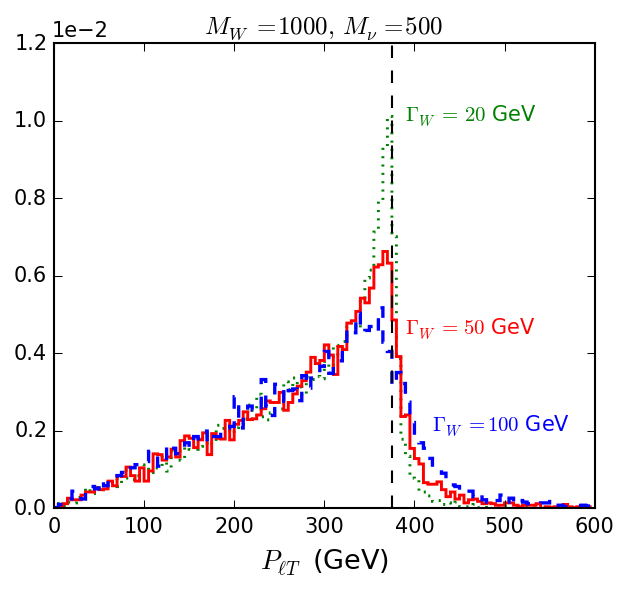}\\
\caption{\label{fig:pt}
The same as Fig.~\ref{fig:ptzero}, but for fixed $M_W=1000$ GeV, and several values of the width $\Gamma_W$ (as indicated in the figure).}
\end{figure}

{\bf Measurement of the width $\Gamma_W$.}~~
The width effects will manifest themselves in two places. First, there will be some smearing of the $P_{\ell T}$ endpoint \cite{Grossman:2011nh} as illustrated in Fig.~\ref{fig:pt}.\footnote{The same effect would be observed in the $M_T$ distribution \cite{Summers:1996yc}.}
%%%%%%%%%%%%% Beginning OF FIGURE ################%%%%%%%%%%%%
%%%%%%%%%%%%% End OF FIGURE %%%%%%%%%%%%%%%%%%%%%%%%%%%%%%%%%
However, the $P_{\ell T}$ distributions resulting from different choices of $M_W$, $M_\nu$, and $\Gamma_W$ will be relatively
similar provided $M_W$ and $M_\nu$ give the same endpoint $\mu$ (following eq.~(\ref{MTendpoint})).
Given this criterion, the distributions will tend to be more similar if the masses and the width satisfy the relation
\begin{eqnarray}
\frac{\Gamma_W}{\Gamma_W^{true}}  = 
\frac{1+\left(M_\nu^{true}/M_W^{true}\right)^2}{1+\left(M_\nu/M_W\right)^2},
\label{GammaW}
\end{eqnarray}
which follows from demanding a similar distribution in the ``endpoints'' obtained from ``off-shell'' $W$ bosons
in the different scenarios.
%%%%%%%%%%%%% Beginning OF FIGURE ################%%%%%%%%%%%%
\begin{figure}[ht]
\includegraphics[width=0.99\columnwidth]{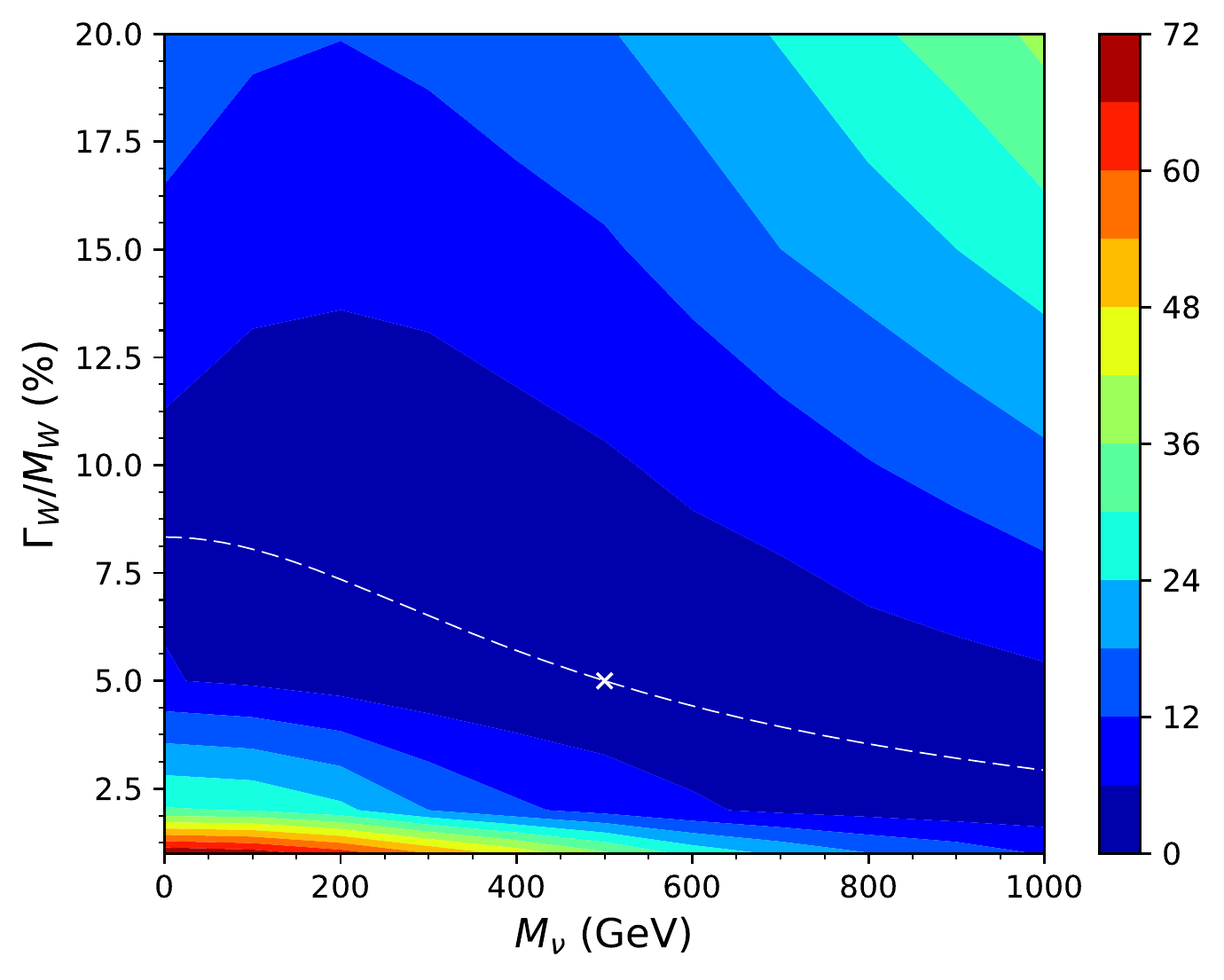}
\caption{\label{fig:GammaM}
Results from a $\chi^2$ fit to the one-dimensional $P_{\ell T}$ distribution for a data sample of 10,000 events. 
The fitted parameters are $M_\nu$ and $\Gamma_W/M_W$,
with $M_{W}$ computed from (\ref{MTendpoint}). The study point ($\times$) has $M_{W}=1000$ GeV, 
$M_{\nu}=500$ GeV, and $\Gamma_W=50$ GeV. The dashed line marks the flat direction (\ref{GammaW}).
The color bar indicates the $\chi^2/d.o.f.$ (we used 100 bins). }
\end{figure}
%%%%%%%%%%%%% End OF FIGURE %%%%%%%%%%%%%%%%%%%%%%%%%%%%%%%%%
Second, the width will also affect the lepton $P_z$ distribution, although to a much smaller extent than the mass scale.
In Fig.~\ref{fig:GammaM}, we examine how well we can simultaneously measure $M_\nu$ and $\Gamma_W$
by determining the $\chi^2$ fit to the one-dimensional $P_{\ell T}$ distribution for the study point with $M_{W} = 1000$ GeV, 
$M_{\nu}=500$ GeV, and $\Gamma_W=50$ GeV, which is indicated on the plot with the $\times$ symbol.
The dashed line marks the relatively flat direction in the $\chi^2$ which is
described by eq.~(\ref{GammaW}).

\begin{figure}[ht]
\includegraphics[width=0.49\columnwidth]{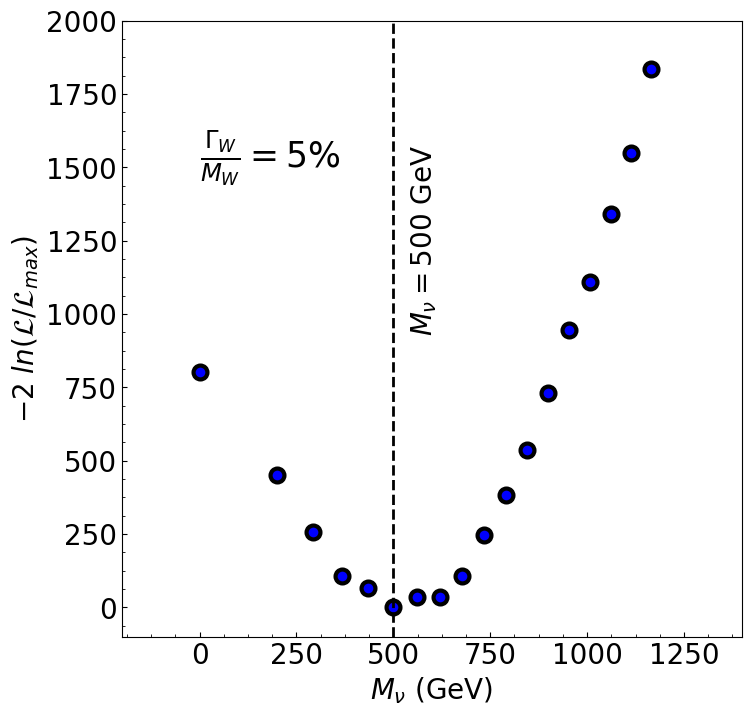}
\includegraphics[width=0.49\columnwidth]{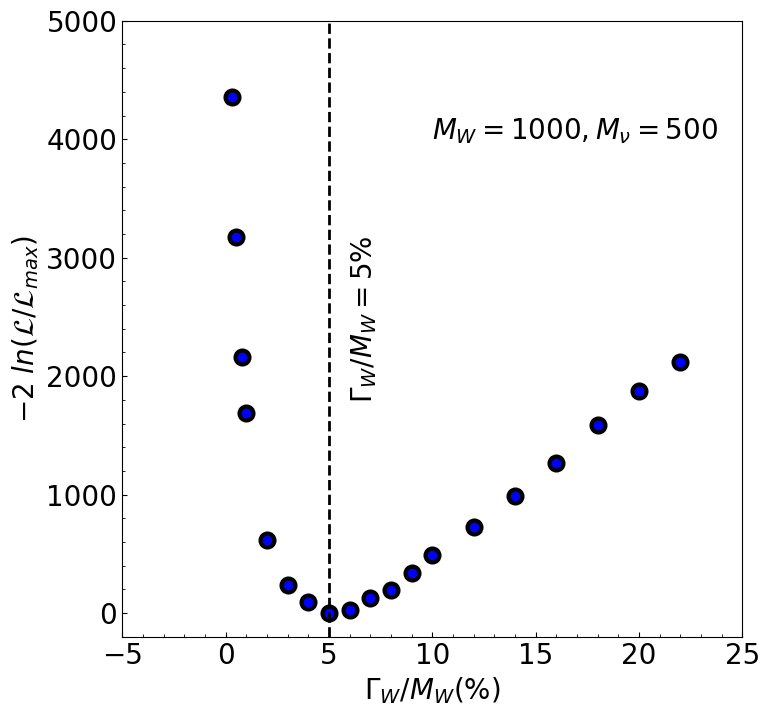}
\caption{\label{fig:GwMw1d}
One-dimensional scan in $M_\nu$ for a fixed width of $5\% \times M_W$ and $M_W$ given by
eq.~(\ref{MWformula}) (left) and a one-dimensional scan in $\Gamma_W$ with  $M_W$ and $M_\nu$ fixed (right),
for the $W^+$ sample, with the likelihood calculated using 10000 events.}
\end{figure}

\begin{figure}[ht]
\includegraphics[width=0.49\columnwidth]{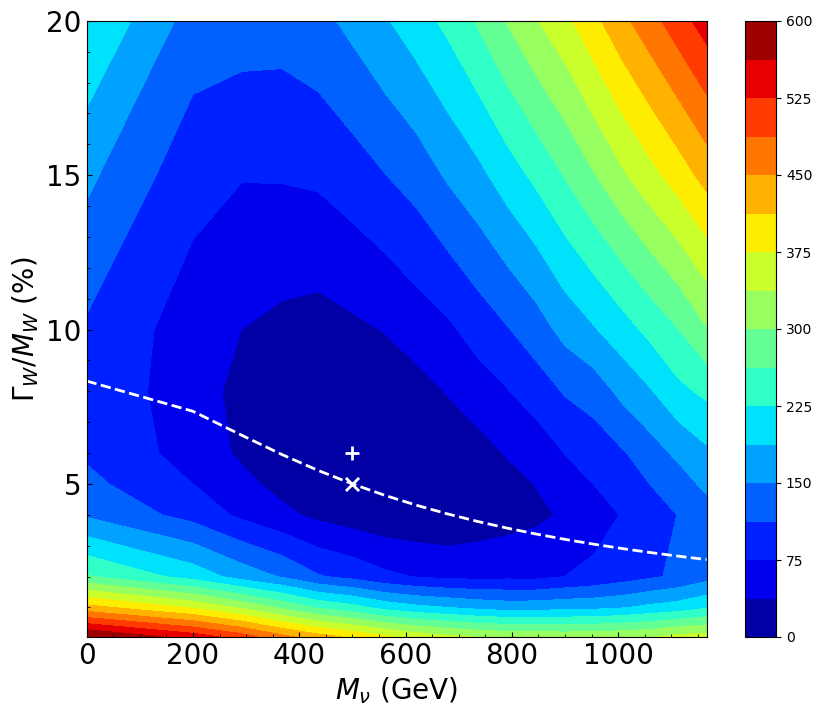}
\includegraphics[width=0.49\columnwidth]{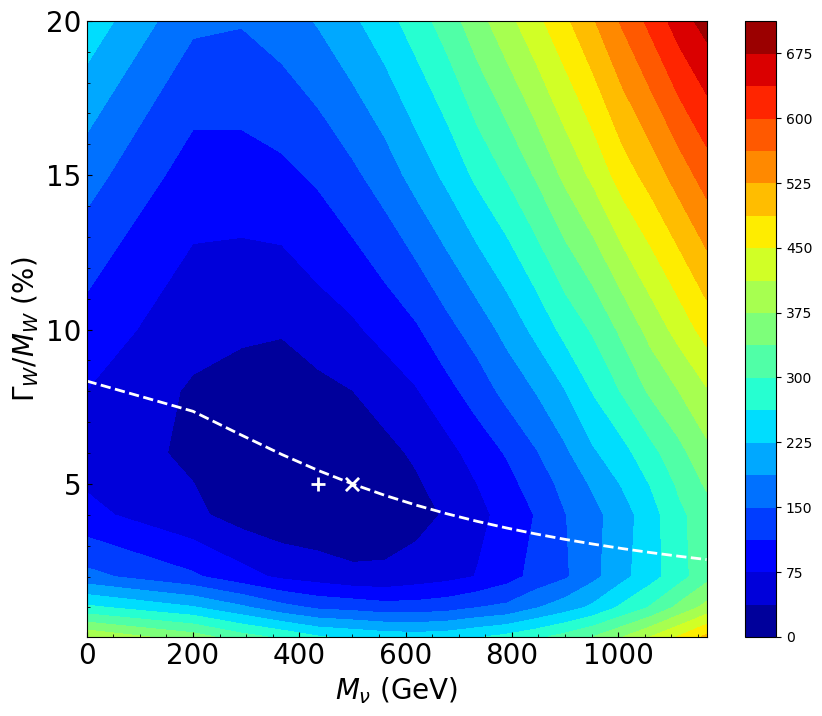}
\caption{\label{fig:GwMw}
A simultaneous measurement of the mass $M_W$ and the width, $\Gamma_W$, 
of the heavy resonance with the MEM for $W^+$ production (left) and $W^-$ production (right). 
The $\times$ ($+$) marks the input values (the result from the fit). The dashed line represents the
relation (\ref{GammaW}).  Contours represent $-2\ln({\cal L}/{\cal L}_{max})$ calculated with 1000 events.
}
\end{figure}

{\bf Simultaneous measurement of the mass scale and the width with the MEM.}~~
We are now ready to measure the mass scale ($M_\nu$) and the width ($\Gamma_W$) with the MEM. 
In Fig.~\ref{fig:GwMw1d} we start by measuring a single parameter, 
for example the mass scale, $M_\nu$ (left panel) or the
width, $\Gamma_W$ (right panel).
%%%%%%%%%%%%% Beginning OF FIGURE ################%%%%%%%%%%%%
%%%%%%%%%%%%% End OF FIGURE %%%%%%%%%%%%%%%%%%%%%%%%%%%%%%%%%
In either case, one has to make an ansatz for the second parameter (the width and the mass scale, respectively).
The ansatz may or may not be correct, which  motivates the {\em simultaneous} measurement of the two parameters with the MEM,
as shown in Fig.~\ref{fig:GwMw}.
%%%%%%%%%%%%% Beginning OF FIGURE ################%%%%%%%%%%%%
%%%%%%%%%%%%% End OF FIGURE %%%%%%%%%%%%%%%%%%%%%%%%%%%%%%%%%
The input values of the parameters were $M_W=1000$ GeV and $M_{\nu}=500$ GeV, which results in $\mu=375$ GeV.
The $W$ width was $5\%$ of its mass, i.e., $\Gamma_W = 50$ GeV. We chose the chirality of the couplings to be left-handed, i.e.,
$g_R^q=g_R^\ell=0$, which was assumed to be known. 
This assumption is harmless, since, as we show in the next section, 
the chirality of the couplings to the $W$ can also be measured analogously with the MEM.

\begin{figure}[ht]
\includegraphics[width=0.49\columnwidth]{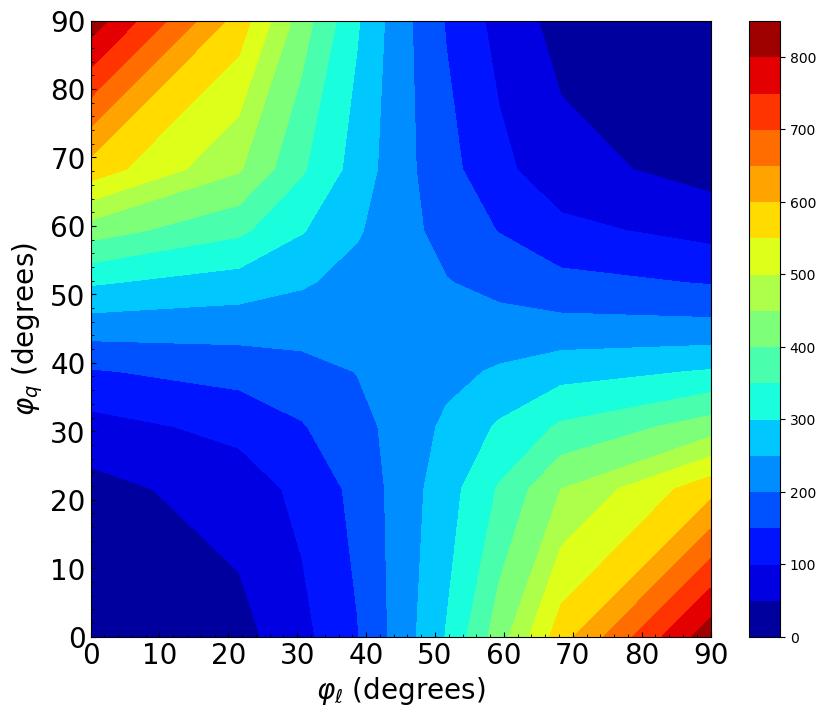}
\includegraphics[width=0.49\columnwidth]{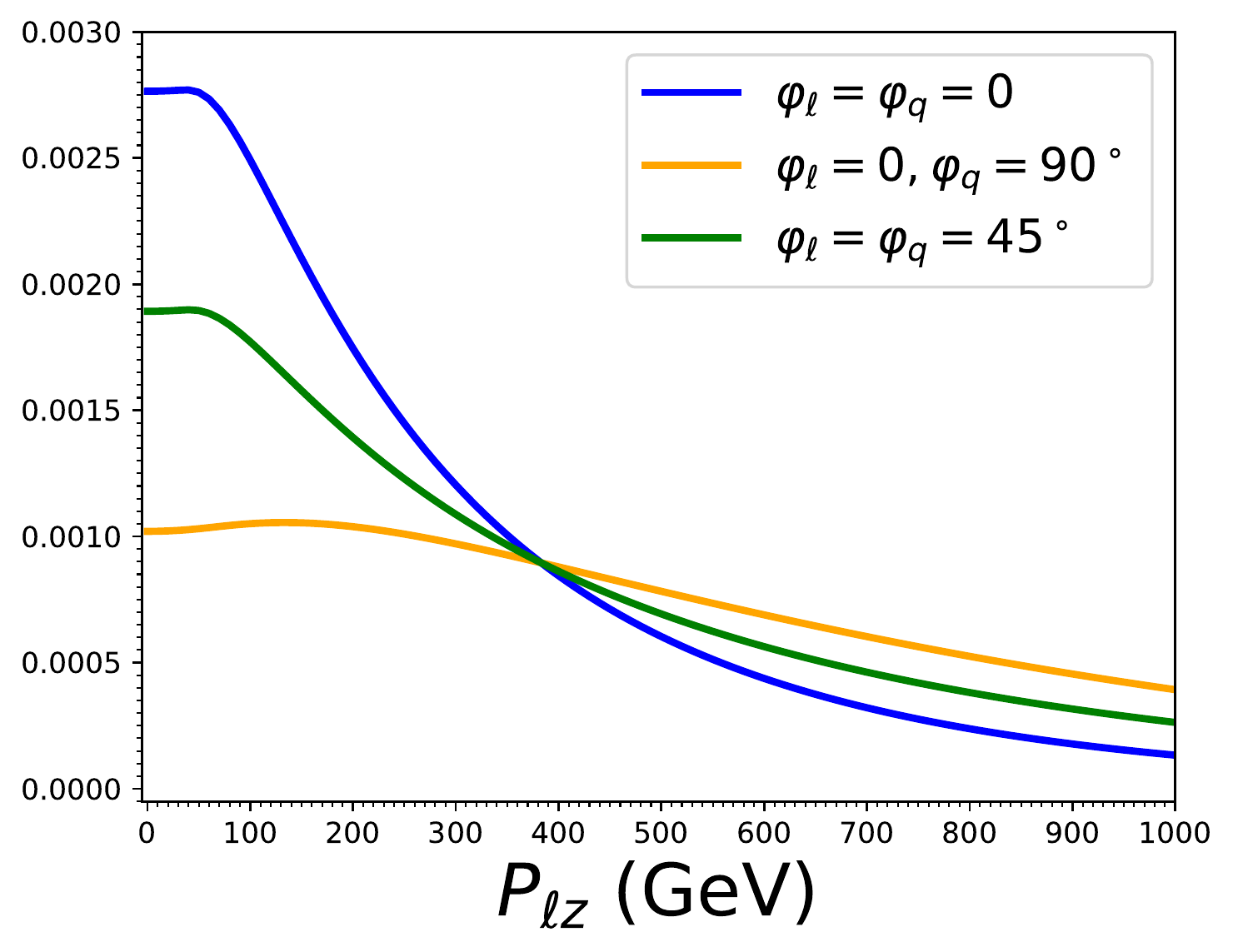}
\caption{\label{fig:chirality}
Left: A fit to the chirality of the quark and lepton couplings to the heavy resonance. The input study point has 
$M_W=1000$ GeV, $\Gamma_W=50$ GeV, $\varphi_\ell=0$, and $\varphi_q=0$.
Contours represent $-2\ln({\cal L}/{\cal L}_{max})$ calculated with 1000 events.
Right: Lepton $P_z$ distributions for different chiralities, obtained from analytic expressions and
{\sc LHAPDF} pdfs~\cite{LHAPDF}  for $M_W=1000$ GeV and $\Gamma_W=0$. }
\end{figure}

{\bf Measurement of the chirality of the couplings.}~~
Having measured the two masses and the width, the only remaining task is to measure the chirality of the couplings\footnote{In practice, 
these measurements will be done simultaneously, see the next section.} 
In analogy to Fig.~\ref{fig:GwMw}, 
Fig.~\ref{fig:chirality} shows a simultaneous extraction of the chirality of the couplings to quarks and leptons,
for fixed $M_{W}=1000$ GeV and $\Gamma_W=50$ GeV (the nominal values measured in Fig.~\ref{fig:GwMw}).
%%%%%%%%%%%%% Beginning OF FIGURE ################%%%%%%%%%%%%
%%%%%%%%%%%%% End OF FIGURE %%%%%%%%%%%%%%%%%%%%%%%%%%%%%%%%%
As one would expect from eq.~(\ref{eq:magic-likelihood}), 
the contour lines in the left plot of Fig.~\ref{fig:chirality} are given by
\beq
\label{eq:contour}
\cos (2 \varphi_\ell) \cos(2 \varphi_q) = {\rm constant}\ \equiv \cos(2\varphi_{\rm rel}),
\eeq
where $\varphi_{\rm rel}$ parametrizes the relative chirality of the two vertices.
Fig.~\ref{fig:chirality} reveals that the chirality of the couplings can be measured very well
(up to the degeneracy described by eq.~(\ref{eq:contour})).
This is because the $P_{\ell z}$ distribution is very sensitive to the chirality\footnote{We note that in the case when the BSM signal of Fig.~\ref{fig:diagrams}(a)
is due to a $W'$ gauge boson decaying to a SM neutrino, the chirality of the couplings can in principle also be determined 
by studying the $W'-W$ interference effects in the $M_T$ distribution \cite{Rizzo:2007xs}.}, as shown in the right panel of 
Fig.~\ref{fig:chirality}, where we plot the $P_{\ell z}$ distribution for the correct mass spectrum and the correct
$\Gamma_W$, but for three different choices of the couplings: $\varphi_q=\varphi_\ell=0^{\degree}$ (blue);
$\varphi_q=\varphi_\ell= 45^{\degree}$ (green); and $\varphi_\ell=0^{\degree}$ and  $\varphi_q=90^{\degree}$ (orange).
The lesson to be learned  from all these exercises so far is that one should not attempt to do mass measurements from 
the shapes of the kinematic distributions unless one is sure about the chiralities of the couplings;
if the chiralities are a priori unknown, then one should fit to the masses, widths and chiralities simultaneously 
\cite{Burns:2008cp}. This ultimate exercise is performed in the next section.

%%%%%%%%%%%%% Beginning OF FIGURE ################%%%%%%%%%%%%
\begin{figure}[t]
\includegraphics[width=0.98\columnwidth]{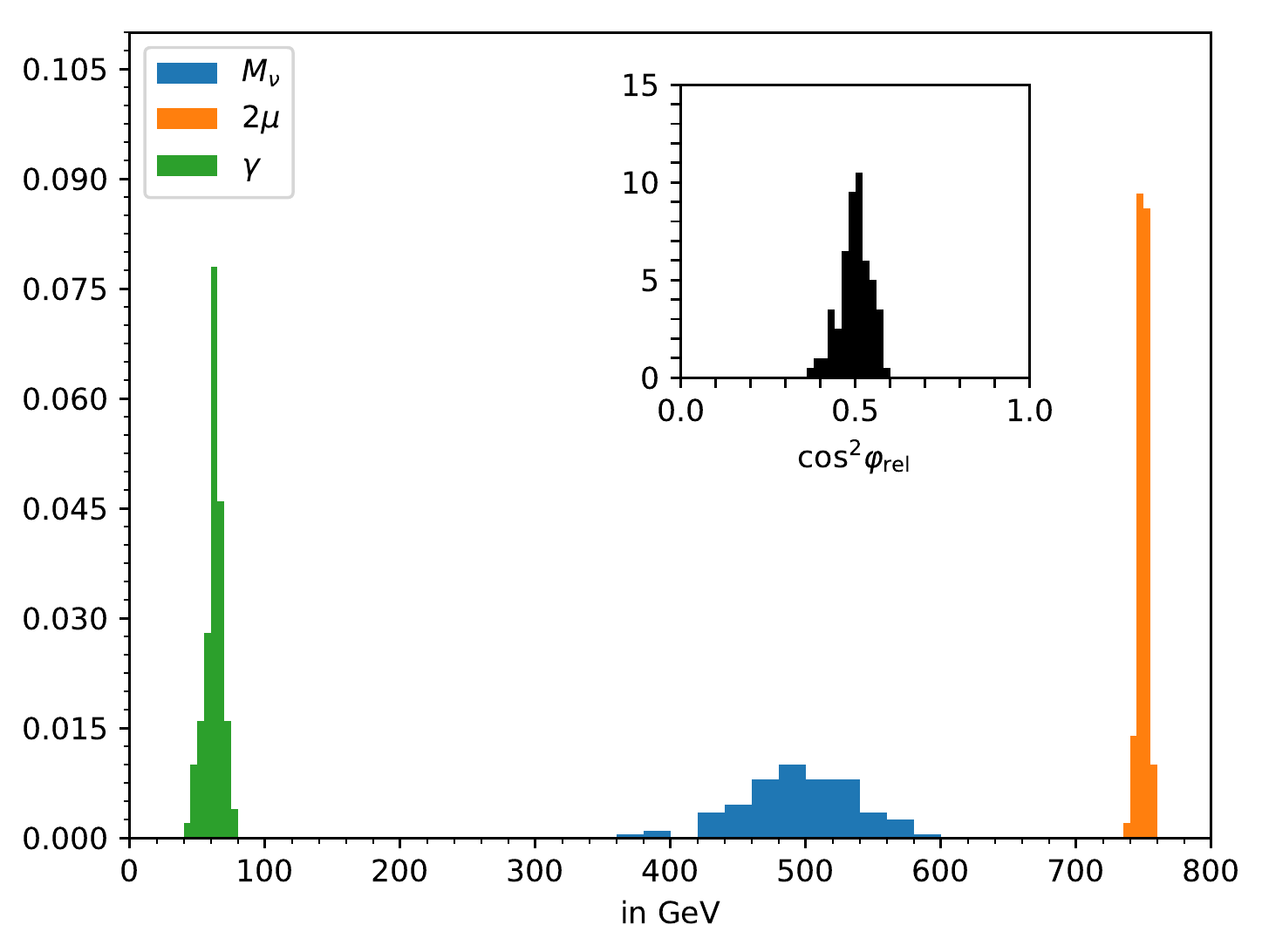}\\
\caption{\label{fig:simul}
Simultaneous measurement of all four parameters: the daughter particle mass $M_\nu$ (blue), 
the parent particle mass parameter $\mu$ defined in (\ref{MTendpoint}) (orange),
the parent particle width parameter $\gamma$ defined in (\ref{eq:gammadef}) (green),
and the relative chirality $\varphi_{\rm rel}$ defined in (\ref{eq:phireldef}) (black).
We show normalized distributions of the measured values for each parameter over 100 samples of 1000 events each, using MEM. 
The input values for our study point were $M_\nu = 500~\mathrm{GeV}$, $M_W = 1000~\mathrm{GeV}$, $\Gamma_W = 50~\mathrm{GeV}$ and $\cos^2{\varphi_\mathrm{rel}} = 0.5$, which
translates into $2 \mu = 750~\mathrm{GeV}$ and $\gamma = 62.5~\mathrm{GeV}$.
}
\end{figure}
%%%%%%%%%%%%% End OF FIGURE %%%%%%%%%%%%%%%%%%%%%%%%%%%%%%%%%

{\bf Simultaneous measurement of all parameters.}~~
After the preliminary exercises shown in the previous sections, we now attempt to simultaneously measure the
relevant parameters (\ref{parameters}) using the MEM.

As already shown in eqs.~(\ref{eq:magic-likelihood}) and (\ref{eq:contour}), we cannot extract the individual
chiralities $\varphi_\ell$ and $\varphi_q$, but only the relative chirality $\varphi_{\rm rel}$
\beq
\cos^2{\varphi_\mathrm{rel}} \equiv \frac{\cos(2\varphi_l)\cos(2\varphi_q) + 1}{2}.
\label{eq:phireldef}
\eeq
As for the remaining three parameters, $M_\nu$, $M_W$, and $\Gamma_W$, they will all share a common source of uncertainty
coming from the overall mass scale, causing their measured values to be highly correlated. In order to 
reduce the covariance between the parameters being measured, we choose to reparametrize them
%mass and width parameters $M_\nu$, $M_W$, and $\Gamma_W$ 
in terms of the daughter mass $M_\nu$, the parent mass parameter $\mu$ from eq.~(\ref{MTendpoint}), and 
the width parameter $\gamma$,
\beq
\gamma = \Gamma_W\left(1+\frac{M_\nu^2}{M_W^2}\right),
\label{eq:gammadef}
\eeq
%Recall that $\mu$ is defined in eq.~\ref{MTendpoint} as
%\beq
%\mu = \frac{M_W^2-M_\nu^2}{2}.
%\eeq
which appears in eq.~(\ref{GammaW}).
From eqs.~(\ref{MTendpoint}) and (\ref{eq:gammadef}) we see that when $M_\nu = 0$,
$2\mu$ and $\gamma$ are identically equal to $M_W$ and $\Gamma_W$ respectively.
With this choice, we expect that $2\mu$ and $\gamma$ will be measured relatively well, while 
the mass scale uncertainty will only be manifested in the determination of $M_\nu$.

These expectations are confirmed in Fig.~\ref{fig:simul}, which shows the results from 
our simultaneous measurement of the four parameters $M_\nu$, $2\mu$, $\gamma$ and $\cos^2\varphi_{\rm rel}$.
For our purpose, we use simulated data samples of 1000 events each, and 
in each case we find the ``measured" values of $M_{\nu}$, $2\mu$, $\gamma$, and $\cos^2\varphi_{\rm rel}$
by maximizing the likelihood (\ref{eq:likelihood}). 
Fig.~\ref{fig:simul} shows the unit-normalized distributions of the measured values of each parameter from 100 such pseudo-experiments.

The sample mean and the standard deviation of the measured values (with the true values quoted in parentheses) are as follows
\begin{align}
M_\nu &= 495 \pm 42~\mathrm{GeV} &(500~\mathrm{GeV})  \label{Mnumeas}\\
2\mu &= 750 \pm 4~\mathrm{GeV} &(750~\mathrm{GeV}) \\
\gamma &= 62.2 \pm 6.5  ~\mathrm{GeV} &(62.5~\mathrm{GeV})  \\
\cos^2{\varphi_\mathrm{rel}} &= 0.499 \pm 0.045 &(0.5) \label{phimeas}
\end{align}
%Note that the sample standard deviations are directly related to the confidence intervals corresponding to the measurements of the parameters made using a single sample of 1000 events. 
We see that the mass difference $\mu$ is very well constrained, while the measurement of the mass scale $M_\nu$ is less precise, as expected from the toy exercises performed in the lead-up to this analysis.

%%%%%%%%%%%%% Beginning OF FIGURE ################%%%%%%%%%%%%
\begin{figure}[t]
\includegraphics[width=0.49\columnwidth]{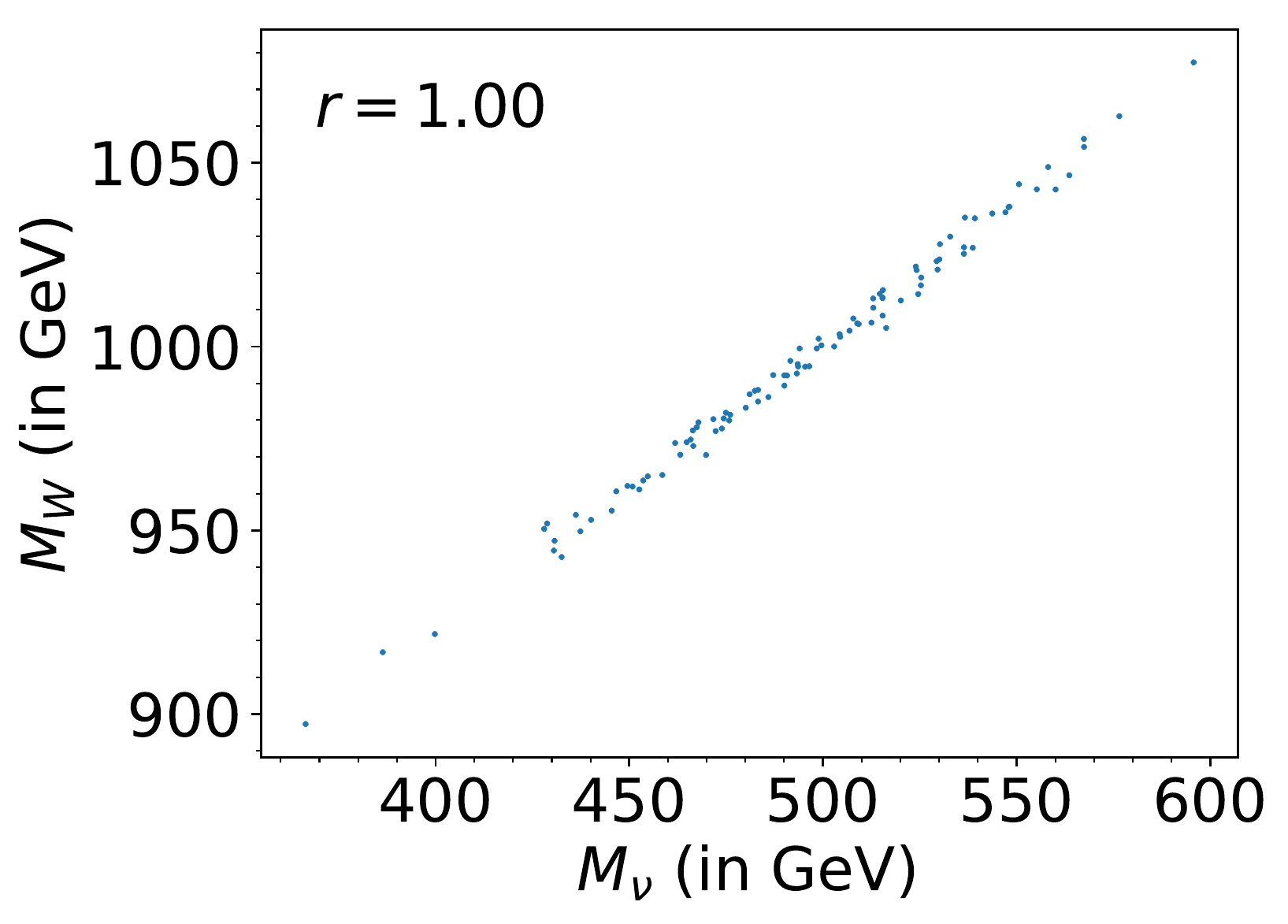}
\includegraphics[width=0.49\columnwidth]{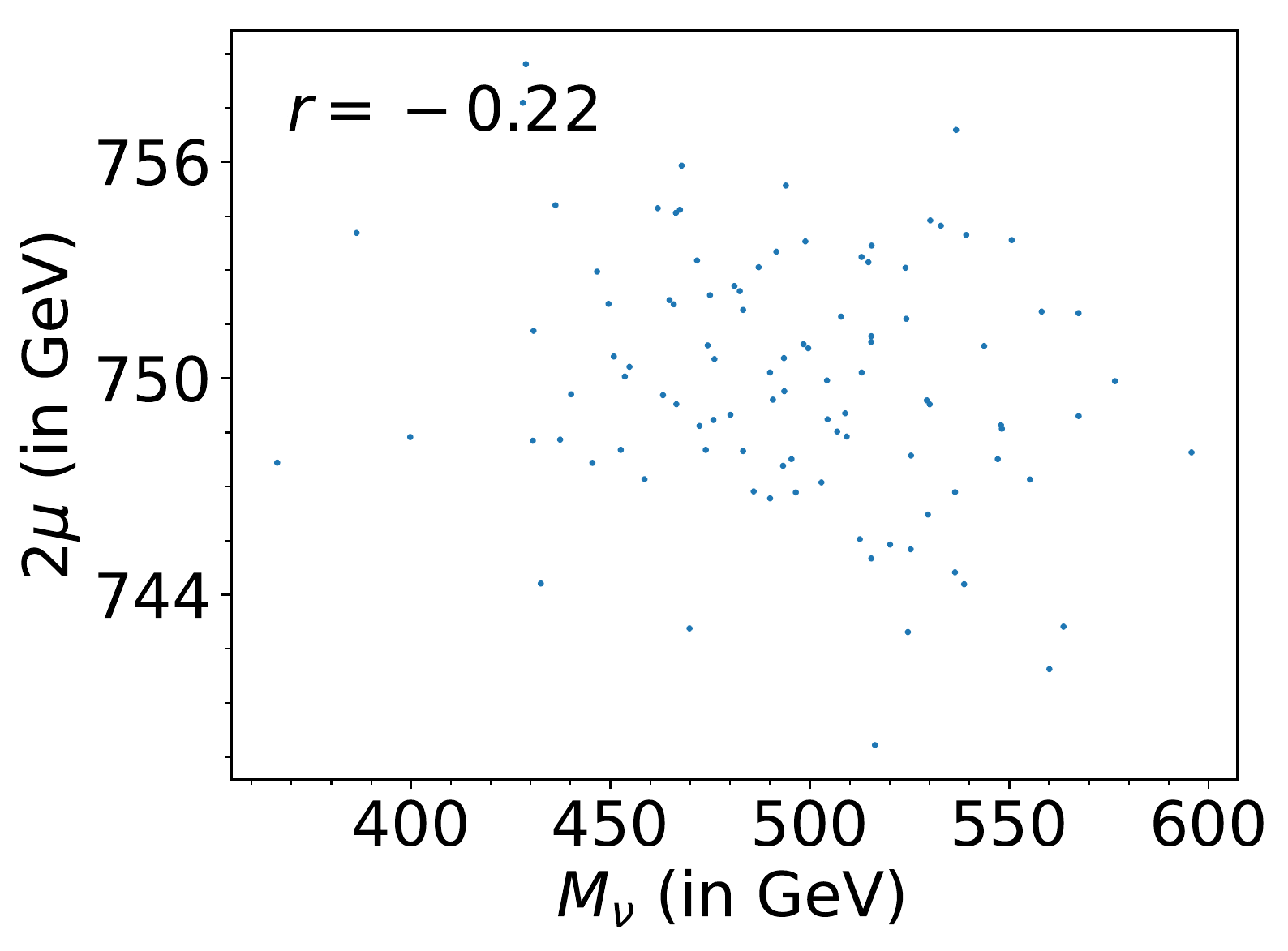}
\\
\includegraphics[width=0.49\columnwidth]{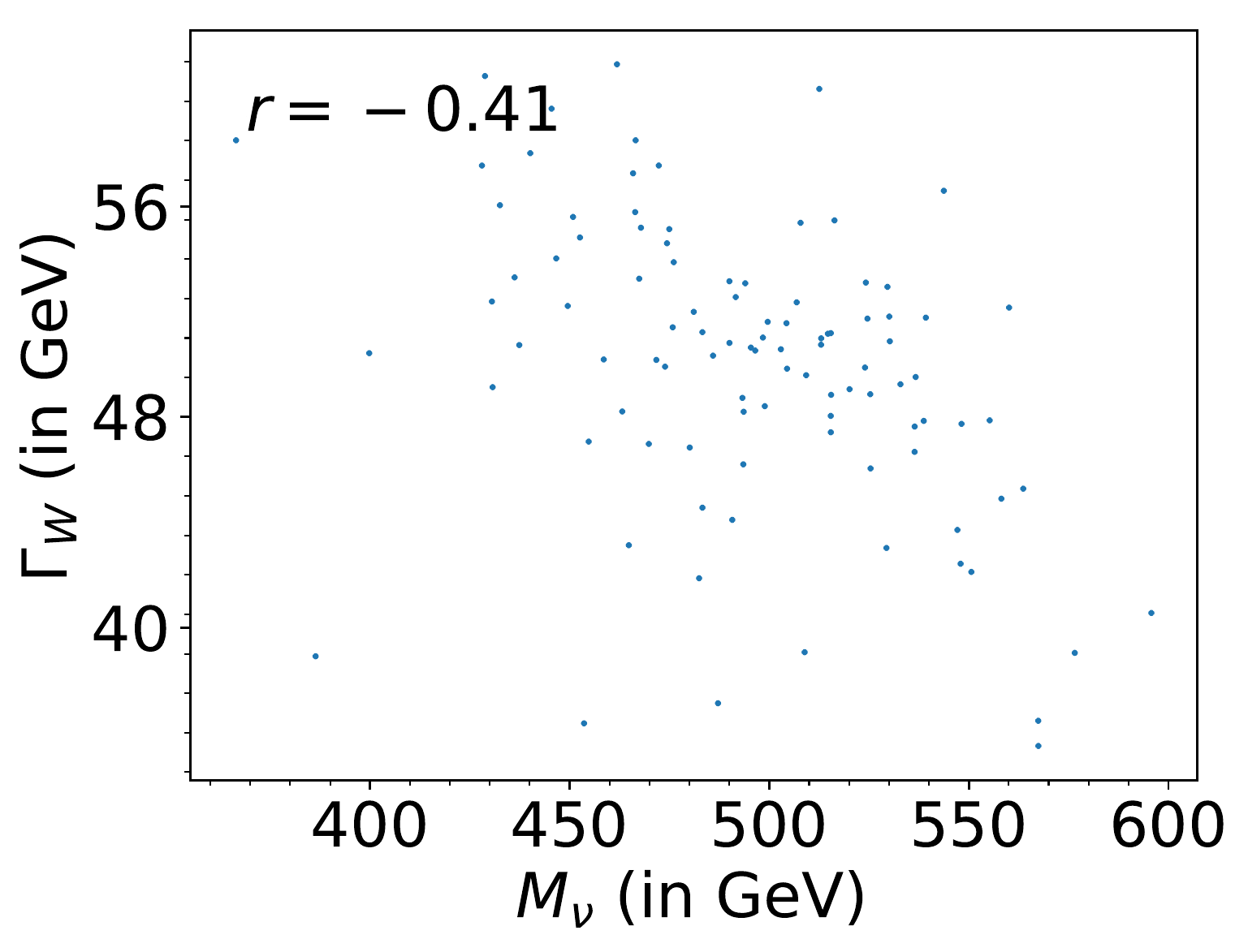}
\includegraphics[width=0.49\columnwidth]{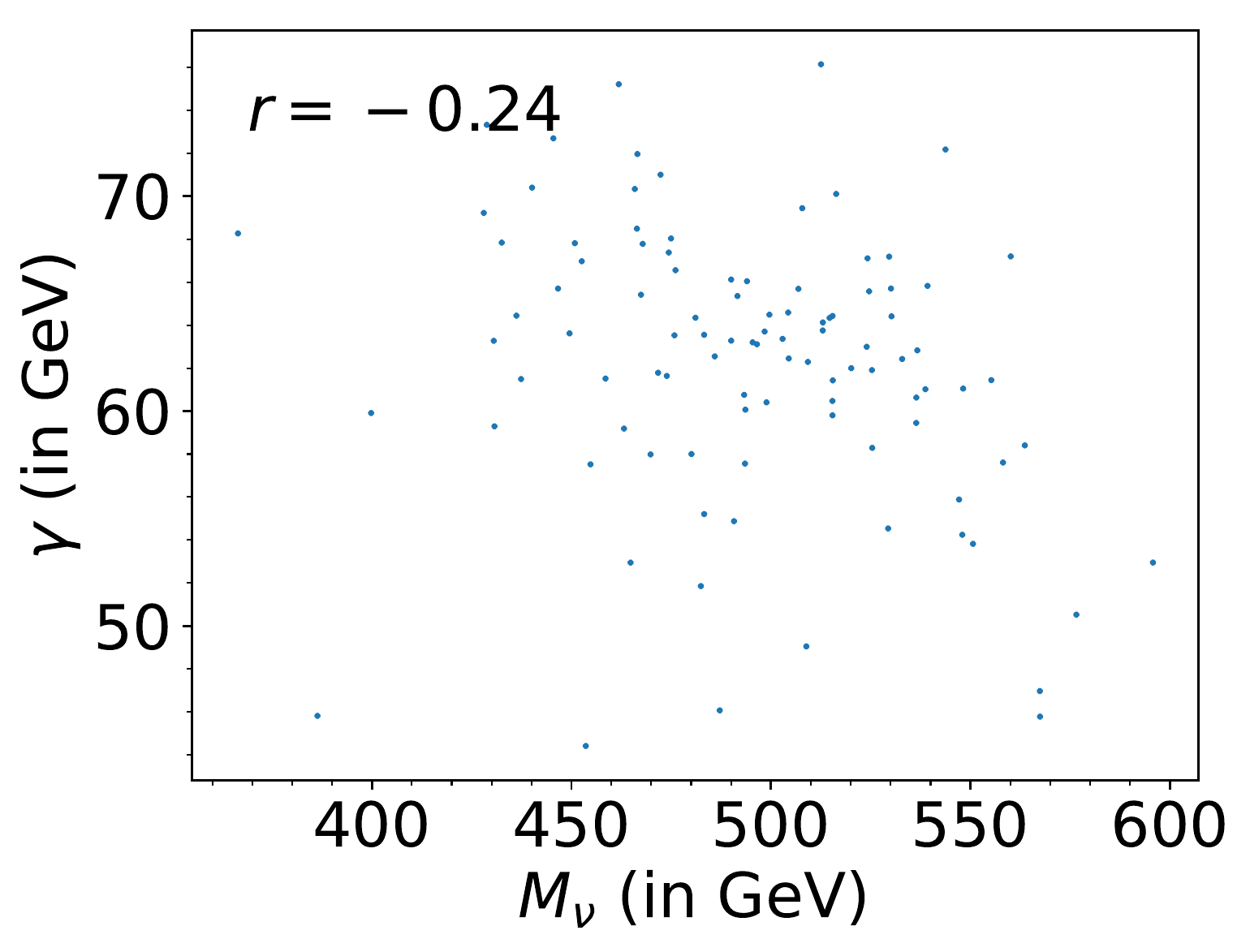}
\caption{\label{fig:correlations}
Correlations among the measured values of the parameters: 
$M_\nu$ and $M_W$ (top left),
$M_\nu$ and $2\mu$ (top right),
$M_\nu$ and $\Gamma_W$ (bottom left) and 
$M_\nu$ and $\gamma$ (bottom right).
}
\end{figure}
%%%%%%%%%%%%% End OF FIGURE %%%%%%%%%%%%%%%%%%%%%%%%%%%%%%%%%

For completeness, we also study the correlations among the different measurements 
(\ref{Mnumeas}-\ref{phimeas}). They are illustrated in Fig.~\ref{fig:correlations}, where we
show scatter plots for the individual measurements in the 100 pseudo-experiments used  in Fig.~\ref{fig:simul}.
The four panels of Fig.~\ref{fig:correlations} focus on certain interesting pairs of variables: $M_\nu$ and $M_W$ (top left),
$M_\nu$ and $2\mu$ (top right),
$M_\nu$ and $\Gamma_W$ (bottom left) and 
$M_\nu$ and $\gamma$ (bottom right).
In the top left corner of each panel in Fig.~\ref{fig:correlations} we show the 
corresponding Pearson correlation coefficient\footnote{The Pearson correlation coefficient between variables $x$ and $y$ is computed as
\beq
r_{xy} = \frac{\sum_i (x_i-\bar{x}) (y_i-\bar{y})}{\sqrt{\sum_i (x_i-\bar{x})^2}\sqrt{ \sum_i(y_i-\bar{y})^2}}.
\eeq
} between the two variables used in the plot.

The top left panel in Fig.~\ref{fig:correlations} demonstrates the significant correlation among the measured values of 
$M_\nu$ and $M_W$. This was the primary motivation for trading $M_W$ for the parameter $\mu$ via
eq.~(\ref{MTendpoint}). Indeed, the top right panel in Fig.~\ref{fig:correlations} confirms that the correlation between
$M_\nu$ and $\mu$ is much milder. Similarly, the correlation between $M_\nu$ and $\Gamma_W$ (bottom left panel)
gets ameliorated once we switch from $\Gamma_W$ to the parameter $\gamma$ defined in (\ref{eq:gammadef}),
see the bottom right panel in Fig.~\ref{fig:correlations}.

{\bf The case of pair production.}~~
Now let us consider pair production, as in the second diagram of Fig.~\ref{fig:diagrams}.
We generate $q \bar{q} \rightarrow W^{+} W^{-} \rightarrow 2\ell+MET$ events using \amc~ at $\sqrt{s}=13$ TeV, again without cuts 
or detector simulation. We set $M_W=1000$ GeV,  $M_\nu=500$ GeV, 
and $\Gamma_W$ to $5 \%$ of the parent mass. For simplicity we keep only the $s$-channel diagram as the $t$- and $u$-channel
diagrams are more model-dependent. Also we do not include 
the $Z$-boson in the $s$-channel diagram because our $W'$-like particle may not participate in the weak interaction.

As in Fig.~\ref{fig:ptzero}, the mass ``splitting" between the $W$ and the $\nu$ can be easily measured, 
this time from the endpoint of the distribution of $M_{T2}$ instead of $M_T$ \cite{Lester:1999tx}. 
The same combination of masses (\ref{MTendpoint}) will be constrained.

\begin{figure}[ht]
\includegraphics[width=0.99\columnwidth]{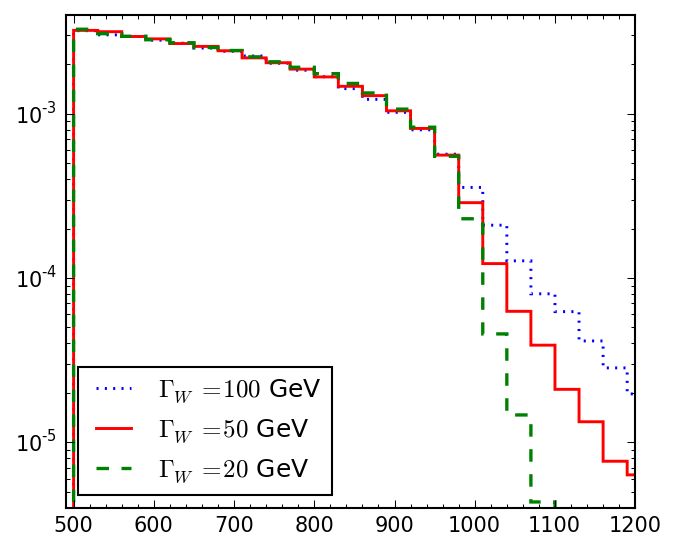}\\
\includegraphics[width=0.99\columnwidth]{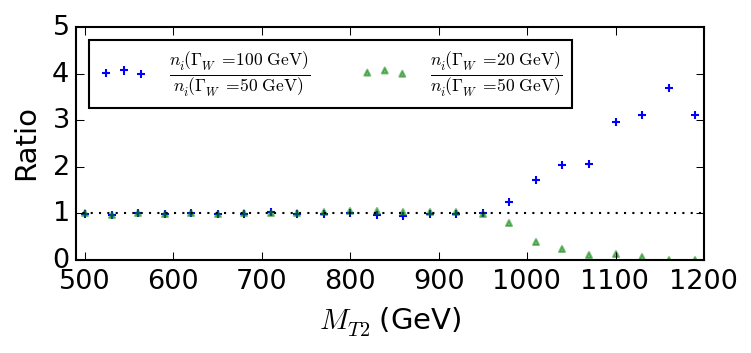}
\caption{\label{fig:MT2}
The same as Fig.~\ref{fig:pt}, but for the case of pair production, where we study the $M_{T2}$ distribution instead of $P_T$.
In the lower panel we show the bin-by-bin ratio of the number of events for different widths, normalized to the case of $\Gamma_W=50$ GeV. }
\end{figure}

In analogy to Fig.~\ref{fig:pt}, in Fig.~\ref{fig:MT2} we show the $M_{T2}$ distribution for several  values of the width, $\Gamma_W$,
illustrating the smearing of the kinematic endpoint. The figure suggests that there is sensitivity to the width, but the measurement is
challenging, since the effect is concentrated in the region near the endpoint, $M_{T2}\sim 950-1200$ GeV. 
We point out that the above measurements can also be performed using the $M_{CT}$ variable \cite{Tovey:2008ui}
--- it similarly has a well defined kinematic endpoint, which will be partially smeared by the width effects.

%%%%%%%%%%%%% Beginning OF FIGURE ################%%%%%%%%%%%%
%%%%%%%%%%%%% End OF FIGURE %%%%%%%%%%%%%%%%%%%%%%%%%%%%%%%%%

\begin{figure}[ht]
\includegraphics[height=0.45\columnwidth]{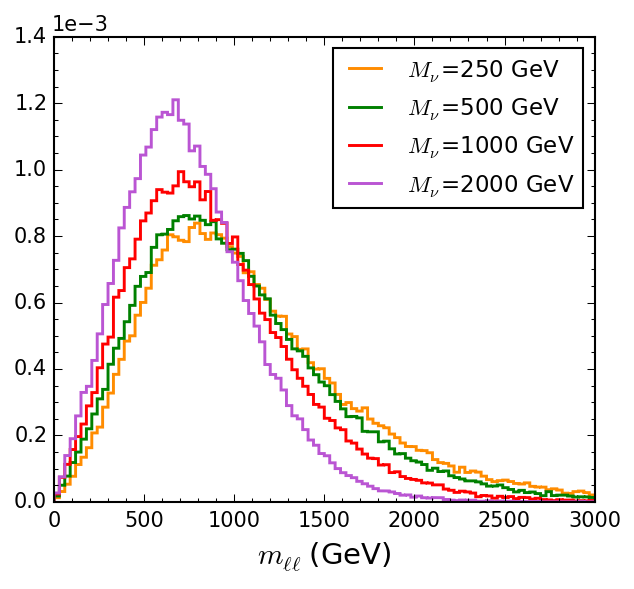}
\includegraphics[height=0.45\columnwidth]{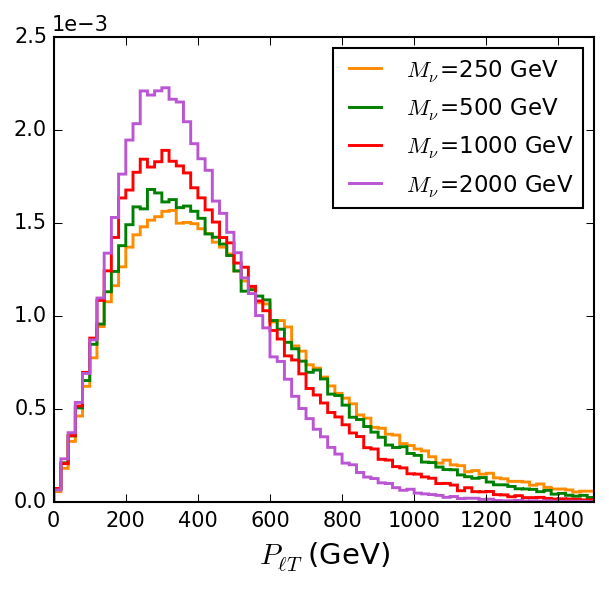}\\
\includegraphics[height=0.45\columnwidth]{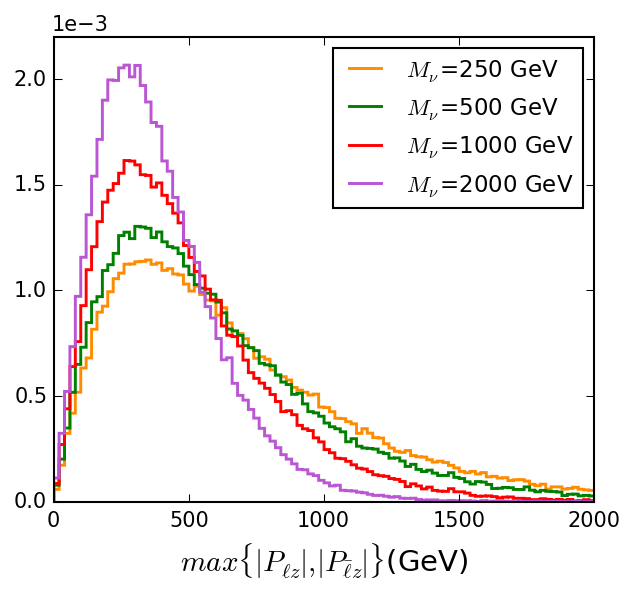}
\includegraphics[height=0.45\columnwidth]{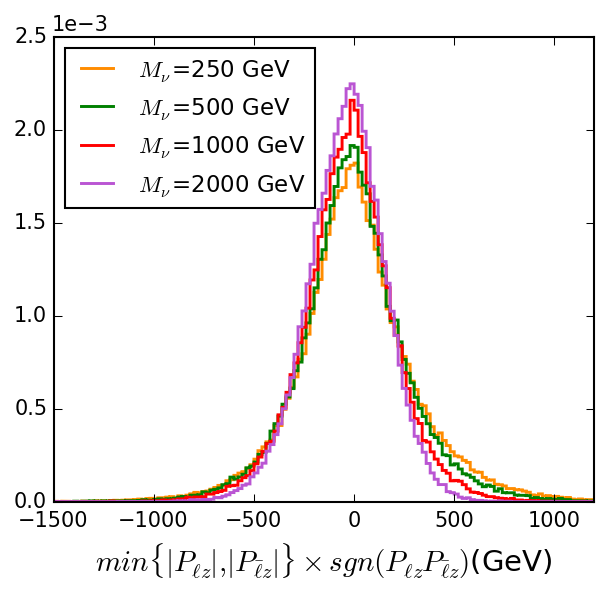}
\caption{\label{fig:mll}
The same as Fig.~\ref{fig:pz}, but for the case of pair production. We showcase several variables whose distributions
are sensitive to the overall mass scale: the dilepton invariant mass, $m_{\ell\ell}$ (upper left),
the transverse lepton momentum, $P_{\ell T}$ (upper right), the larger of the two longitudinal lepton momenta (in absolute value),  
$\max(|P_{\ell z}|,|P_{\bar{\ell} z}|)$ (lower left), and the other longitudinal lepton momentum with its sign chosen as
${\rm sgn}(P_{\ell z}P_{\bar{\ell} z})\min(|P_{\ell z}|,|P_{\bar{\ell} z}|)$ (lower right). }
\end{figure}

As in the case of single production, for the measurement of the mass scale one cannot rely on endpoint 
measurements alone and needs to utilize the shapes of the relevant kinematic distributions.
Fig.~\ref{fig:mll} depicts several variables whose distributions show sensitivity to the overall mass scale.
%
%%%%%%%%%%%%% Beginning OF FIGURE ################%%%%%%%%%%%%
%%%%%%%%%%%%% End OF FIGURE %%%%%%%%%%%%%%%%%%%%%%%%%%%%%%%%%
%
As in Fig.~\ref{fig:pz}, we only consider mass spectra which obey the relation (\ref{MTendpoint}) and therefore
satisfy the measured $M_{T2}$ (or $M_{CT}$) kinematic endpoint. More specifically, we vary the mass, $M_\nu$, of the 
invisible particle as shown in each panel, then choose the parent mass, $M_W$, from eq.~(\ref{MWformula}).
The main effect of the mass scale is to provide a different boost of the parent particles: lighter (heavier) $W$s will
be produced with a higher (lower) boost. While the parent boost itself is unobservable, its effects are 
reflected in the kinematic distributions of the visible decay products.
For example, when the $W$s are highly boosted in the transverse plane, we would expect the leptons to be back to back.
Any time the $W$s are highly boosted in the CM frame (whether in the transverse plane or along the beam axis)
we expect a larger invariant mass of the visible leptons, as well as higher (on average) values of their 
transverse and longitudinal momenta. 
These expectations are confirmed by Fig.~\ref{fig:mll}, in which we show distributions of the
dilepton invariant mass, $m_{\ell\ell}$ (upper left panel), the transverse lepton momentum, $P_{\ell T}$ (upper right panel), 
and the longitudinal lepton momenta (lower two panels). We note that the boost effect is seen better in the 
distribution of the {\em larger} of the two longitudinal lepton momenta (in absolute value),  
$\max(|P_{\ell z}|,|P_{\bar{\ell} z}|)$, which is shown in the lower left panel. 
The other longitudinal lepton momentum, $\min(|P_{\ell z}|,|P_{\bar{\ell} z}|)$, 
is then plotted in the lower right panel, with the sign chosen so that it is 
positive (negative) when the two longitudinal lepton momenta have equal (opposite) signs.

Fig.~\ref{fig:mll} demonstrates that the kinematic distributions of the visible particles do, in principle, contain 
information about the mass scale, which can then be extracted from a fit to these (one-dimensional) distributions, 
as was done in Fig.~\ref{fig:template}. However, the MEM will have better sensitivity, as it takes into account the {\em correlations}
among the different variables.

{\bf Parameter measurements in the case of pair production with the MEM.}~~
In analogy to single production, we now apply the MEM to measure simultaneously the mass scale, the width
(Fig.~\ref{fig:GwMwPairProd}), and the chirality of the lepton couplings (Fig.~\ref{fig:chirality2}).
As expected, the MEM is quite successful in determining all the parameters in (\ref{parameters}).
%
%%%%%%%%%%%%% Beginning OF FIGURE ################%%%%%%%%%%%%
%
\begin{figure}[ht]
\includegraphics[width=0.99\columnwidth]{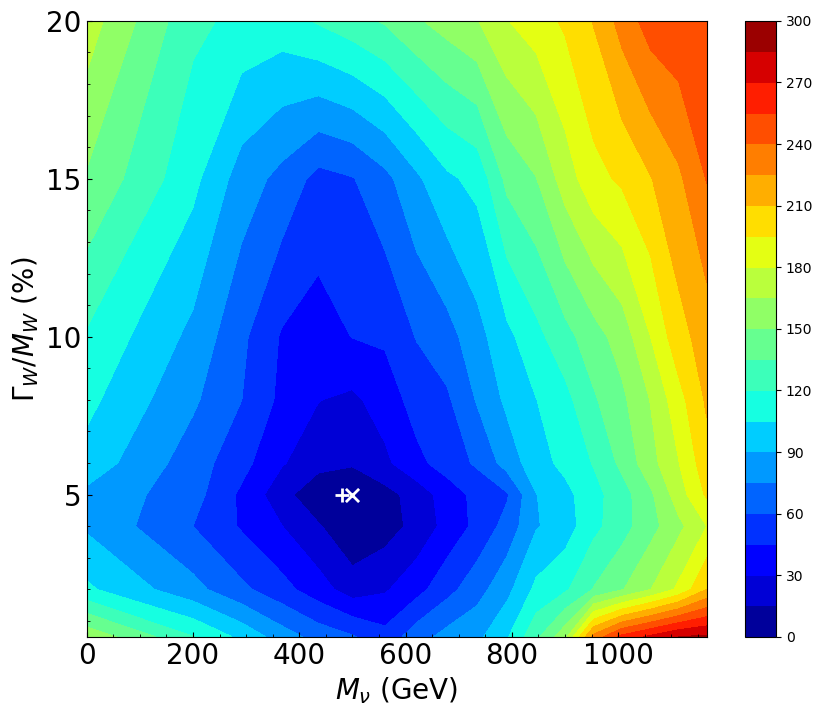}
%\vskip 5cm
\caption{\label{fig:GwMwPairProd}
The same as the left panel in Fig.~\ref{fig:GwMw}, but for pair-production, i.e., the second diagram in Fig.~\ref{fig:diagrams}.
%Contours represent the negative log likelihood from 500 events.  
Contours represent $-2\ln({\cal L}/{\cal L}_{max})$ calculated with 500 events.
The $\times$ ($+$) marks the input values (the result from the fit).
}
\end{figure}
%
%%%%%%%%%%%%% End OF FIGURE %%%%%%%%%%%%%%%%%%%%%%%%%%%%%%%%%
%
In particular, the $+$ symbol in Fig.~\ref{fig:GwMwPairProd} denotes the result from our fit, 
which is close to the input parameter values (marked with $\times$).
%
%%%%%%%%%%%%% Beginning OF FIGURE ################%%%%%%%%%%%%
\begin{figure}[ht]
\includegraphics[width=0.49\columnwidth]{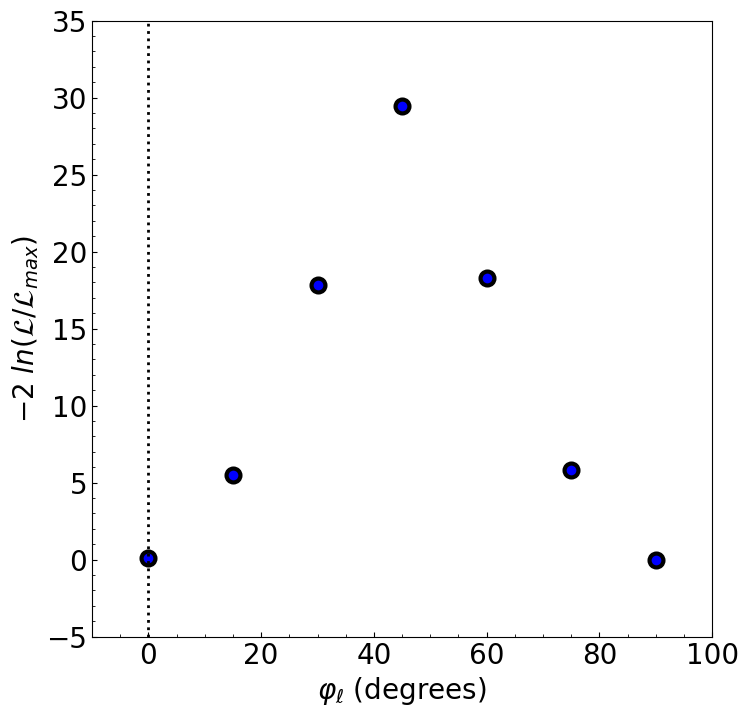}
\includegraphics[width=0.49\columnwidth]{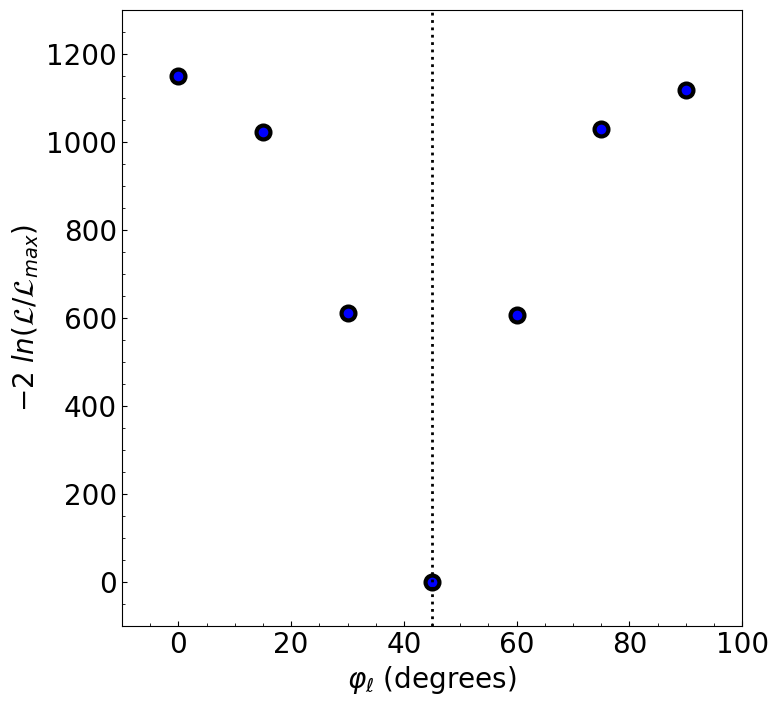}
\caption{\label{fig:chirality2}
A fit to the chirality of the lepton couplings in the case of pair production as in the second diagram of Fig.~\ref{fig:diagrams}
using 600 events.
In the left panel, the couplings were chosen to be purely chiral, $\varphi_\ell=0$,
while in the right panel they were vector-like: $\varphi_\ell=45^\circ$. }
\end{figure}
%%%%%%%%%%%%% End OF FIGURE %%%%%%%%%%%%%%%%%%%%%%%%%%%%%%%%%
%
In Fig.~\ref{fig:chirality2}, a minimum of the negative log-likelihood distribution is always found at
the true input value for the chirality (marked with a vertical dotted line).
Note that in general, the chirality is determined only up to a two-fold ambiguity, $\varphi_\ell \to \pi/2 - \varphi_\ell$,
reflecting the symmetry of the underlying squared amplitude.

The results displayed in Figs.~\ref{fig:GwMwPairProd} and \ref{fig:chirality2} demonstrate the 
power of the MEM for parameter measurements in the challenging event 
topology of Fig.~\ref{fig:diagrams}(b),
thus generalizing and strengthening the conclusions from the previous studies performed 
in Refs.~\cite{Alwall:2009sv,Artoisenet:2010cn,Chen:2010ek}.

{\bf Conclusions.}~~We have presented methods to measure masses, widths, and couplings,
and, in particular, shown that (a) all of these parameters can be simultaneously measured using the 
MEM and (b) that the physics which gives sensitivity to each of these parameters is more transparent than it would be with many MVAs.
While we have focused on the case of $W$s/ $W^\prime$s, the approach and conclusions 
can be readily generalized to nearly any SM or BSM scenario.
We therefore look forward to the utilization of such methods during the continued successful operation of the LHC.

{\bf Acknowledgements.}
K.M. would like to thank his CMS colleagues for useful
discussions. AB and DD thank O. Mattelaer for technical help and
advice. Work supported in part by U.S. Department of Energy Grants DE-SC0010296 and DE-SC0010504. 
AB is grateful for the hospitality of the high energy theory group at the University of Florida
 and acknowledges support from Fulbright and Colciencias.
DD acknowledges support from the University of Florida Informatics Institute in the form of a
Graduate Student Fellowship. 
This work was performed in part at the Aspen Center for Physics,
 which is supported by National Science Foundation grant PHY-1607611.

\end{document}